\documentclass[12pt]{article}
\usepackage{geometry}
\usepackage{amssymb}
\usepackage{amsmath}
\usepackage{bbm}
\usepackage{MnSymbol}
\usepackage{hyperref}
\usepackage{braket}
\usepackage{tikz-cd}
\usepackage{cite}
\usepackage{xcolor}
\usepackage{comment}
\usepackage{subcaption,tikz}
\usepackage{float}
\usepackage{mathtools}
\usepackage{tikz-3dplot}
\usetikzlibrary{arrows,decorations.markings,decorations.pathreplacing,decorations.shapes,decorations.pathmorphing}

\newcommand{\be}{\begin{equation}}
	\newcommand{\ee}{\end{equation}}
\newcommand{\bea}{\begin{eqnarray}}
	\newcommand{\eea}{\end{eqnarray}}

\newcommand{\al}{\alpha}

\newcommand{\mcZ}{\mathcal{Z}}

\newcommand{\C}{\mathbb{C}}

\newcommand{\Z}{\mathbb{Z}}

\newcommand{\lp}{\left(}
\newcommand{\rp}{\right)}

\def\sqr#1#2{{\vcenter{\vbox{\hrule height.#2pt
				\hbox{\vrule width.#2pt height#1pt \kern#1pt
					\vrule width.#2pt}\hrule height.#2pt}}}}

\def\square
{\mathop{\mathchoice{\sqr{12}{15}}{\sqr{9}{12}}
		{\sqr{6.3}{9}}{\sqr{4.5}{9}}}}

\tikzset{
	partial ellipse/.style args={#1:#2:#3}{
		insert path={+ (#1:#3) arc (#1:#2:#3)}
	}
}

\numberwithin{equation}{section}

\begin{document}
	\begin{center}
		
		{\large\bf Anomaly resolution by non-invertible symmetries}

$\,$

Alonso Perez-Lona$^1$, Daniel Robbins$^2$, Subham Roy$^2$, Eric Sharpe$^1$, \\
Thomas Vandermeulen, Xingyang Yu$^1$

        \vspace*{0.1in}
        
        \begin{tabular}{cc}
                {\begin{tabular}{l}
                $^1$ Department of Physics MC 0435\\
                                850 West Campus Drive\\
                                Virginia Tech\\
                                Blacksburg, VA  24061 \end{tabular}}
                                 &
                {\begin{tabular}{l}
                $^2$ Department of Physics\\
                                University at Albany\\
                                Albany, NY 12222 \end{tabular}}
        \end{tabular}
                        
        \vspace*{0.2in}

{\tt aperezl@vt.edu}, 
{\tt dgrobbins@albany.edu},
{\tt sroy3@albany.edu},
{\tt ersharpe@vt.edu},
{\tt tvand@alum.mit.edu},
{\tt xingyangy@vt.edu}
    
	\end{center}

$\,$

In this paper we generalize previous results on anomaly resolution to noninvertible symmetries. 
Briefly, given a global symmetry $G$ of some theory with a 't Hooft anomaly rendering it ungaugeable, the idea of anomaly resolution is to extend $G$ to a larger anomaly-free symmetry of the same theory with a trivially-acting kernel.  In previous work, several of the coauthors  demonstrated that in two-dimensional theories, by virtue of decomposition, gauging the larger symmetry is equivalent to a disjoint union of theories in which a nonanomalous subgroup of $G$ is gauged.  In this paper, we consider examples in which the larger symmetry is not a group, but instead a noninvertible symmetry defined by some fusion category.  In principle the same ideas apply to the case that $G$ itself is noninvertible.  We discuss the construction of larger symmetries using both SymTFT methods as well as algebraically via (quasi-)Hopf algebras.

$\,$

\begin{flushleft}
    April 2025
\end{flushleft}

 \newpage
 \tableofcontents
 \newpage

\section{Introduction}

For ordinary groups, the idea of anomaly resolution \cite{Wang:2017loc,Tachikawa_2020} is as follows.  Given a theory ${\cal T}$ with some global symmetry $G$ with a 't Hooft anomaly, one replaces $G$ by a larger symmetry group $\Gamma$ with trivially-acting kernel $K$, acting on the same theory ${\cal T}$, related by
\begin{equation}
    1 \: \longrightarrow \: K \: \longrightarrow \: \Gamma \: \longrightarrow \: G \: \longrightarrow \: 1,
\end{equation}
such that the action of $\Gamma$ is anomaly-free and hence gaugeable.  Previous work by several of the authors \cite{Robbins_2021,Robbins:2021xce,Robbins:2021ibx,Robbins:2022wlr} has 
argued that by virtue of decomposition \cite{Hellerman:2006zs,Sharpe:2022ene}, gauging $\Gamma$ in two-dimensional theories is equivalent to a disjoint union of theories in which one gauges anomaly-free subgroups $H_i \subset G$.  Schematically,
\begin{equation}
    \left[ {\cal T} / \Gamma \right] \: = \: \coprod_i \left[ {\cal T} / H_i \right]_{ \omega_i },
\end{equation}
where $\omega_i \in H^2(H_i,U(1))$ are choices of discrete torsion.  This relationship was given a completely systematic, predictive, expression in \cite{Robbins:2021xce}.

In this paper, we discuss examples of analogous resolutions in which the groups above are replaced by noninvertible symmetries, defined by fusion categories.  We discuss constructions from both the perspective of SymTFTs and also algebraically, as exercises in exact sequences of
(quasi-)Hopf algebras.  We leave systematic predictions valid for all cases, analogous to those
for ordinary groups in \cite{Robbins:2021xce}, for future work.

We begin in section~\ref{sect:rev} by reviewing anomaly resolution for ordinary groups
and the role of decomposition \cite{Wang:2017loc,Tachikawa_2020,Robbins_2021,Robbins:2021xce,Robbins:2021ibx,Robbins:2022wlr}.
In section~\ref{sect:genl} we describe a more general procedure, in terms of (possibly) noninvertible symmetries, in the language of SymTFTs.  In section~\ref{sect:genl:quasihopf} we describe what is morally the same procedure from an algebraic perspective, in terms of monoidal functors between (quasi-)Hopf algebras.  In section~\ref{sect:reln} we discuss how these two different-looking procedures are equivalent to one another.

Next, we turn to examples.
In section~\ref{sect:redo} we discuss how anomaly-resolution procedure for ordinary groups can be understood in the language of SymTFTs and quasi-Hopf algebras, as a special case.
In section~\ref{sect:exs} we turn to more interesting examples, focusing especially on examples in which an anomalous ${\mathbb Z}_2$ is resolved using a noninvertible symmetry group.

Finally, we collect some technical results in several appendices.
Appendix~\ref{app:red} discusses the ideas of missing charges and their relation with trivially acting symmetries. In addition we also give an example computation of reduced topological order;
appendix~\ref{app:condensable} defines condensable algebras and distinguishes them from gaugeable algebras; appendix~\ref{app:exactseq} defines exact sequences of tensor categories.

\section{Review of anomaly resolution in groups}   \label{sect:rev}

In this section we will briefly review results on anomaly resolution for
ordinary groups as discussed in \cite{Wang:2017loc,Tachikawa_2020,Robbins_2021,Robbins:2021xce,Robbins:2021ibx,Robbins:2022wlr}.
In later sections, we will describe how these results can be generalized to the case of noninvertible symmetries.

\subsection{General picture}
\label{subsec:GroupGeneralPicture}

Begin with a finite group $G$ acting on a two-dimensional theory, with
a 't Hooft anomaly $[\alpha] \in H^3(G,U(1))$ with representative $\alpha\in Z^3(G,U(1))$.  To resolve this anomaly, we look for a larger group $\Gamma$ that projects to $G$ with kernel some abelian group $K$,
\be  \label{eq:ext1}
1\longrightarrow K\stackrel{i}{\longrightarrow} \Gamma\stackrel{\pi}{\longrightarrow} G\longrightarrow 1,
\ee
such that the pullback $\pi^\ast[\alpha]=[\pi^\ast\alpha]\in H^3(\Gamma,U(1))$ is trivial, i.e.~we can find a 2-cochain $j\in C^2(\Gamma,U(1))$ such that $dj=\pi^\ast\alpha$, or explicitly
\begin{equation}
\label{eq:TrvializationOfAlpha}
    \frac{j(\gamma_2,\gamma_3)j(\gamma_1,\gamma_2\gamma_3)}{j(\gamma_1\gamma_2,\gamma_3)j(\gamma_1,\gamma_2)}=\al(\pi(\gamma_1),\pi(\gamma_2),\pi(\gamma_3)),\qquad\forall\gamma_1,\gamma_2,\gamma_3\in\Gamma.
\end{equation}
In this situation $\Gamma$ is anomaly free and it (or any of its subgroups) can be gauged.

The Lyndon-Hochschild-Serre spectral sequence can be used to connect the group cohomology of $\Gamma$ to the cohomology groups of $G$ and $K$.  This can be used to find constructions of trivializations.  Explicitly, pick a section $s:G\rightarrow\Gamma$ of $\pi$, i.e.~a map such that $\pi(s(g))=g$ for all $g\in G$.  $s$ will not, in general, be a homomorphism (only if the extension is split will a homomorphism section exist).  Without loss of generality, we will restrict to sections that send the identity element of $G$ to the identity element of $\Gamma$, $s(e_G)=e_\Gamma$.  Using the section we can construct the extension class $[\bar{c}]\in H^2(G,K_{ab})$, where $K_{ab}=K/[K,K]$ is the abelianization of $K$.  Note that there is an action of $G$ on $K_{ab}$ given by $g\cdot\bar{k}=\overline{s(g)ks(g)^{-1}}$.  To specify the extension class we first construct a map $c:G\times G\rightarrow K$ as
\begin{equation}
    c(g_1,g_2)=s(g_1)s(g_2)s(g_1g_2)^{-1}.
\end{equation}
We then take the image $\bar{c}$ of this map under the quotient $K\rightarrow K_{ab}$.  One can check that $\bar{c}$ is in fact a cocycle in $Z^2(G,K_{ab})$ with the action of $G$ on $K_{ab}$ described above.  Different choices of section $s$ will lead to cocycles $\bar{c}$ that differ by an exact cocycle, but the extension class $[\bar{c}]\in H^2(G,K_{ab})$ will be independent of the choice of section and is determined only by the short exact sequence \eqref{eq:ext1}.

The Lyndon-Hochschild-Serre spectral sequence gives us a map 
\begin{equation}
    d_2: \: H^1(G,H^1(k,U(1))) \: \longrightarrow \: H^3(G,U(1)).
\end{equation}
Indeed, we can think of elements $\beta\in C^1(G,H^1(K,U(1)))$ as maps from $G\times K$ to $U(1)$ that are homomorphisms in their second argument, i.e.~
\begin{equation}
    \beta(g,k_1)\beta(g,k_2)=\beta(g,k_1k_2).
\end{equation}
The cocycle condition is that they are a crossed homomorphism in their first argument, i.e.~$\beta\in Z^1(G,H^1(K,U(1)))$ if and only if 
\begin{equation}
    \beta(g_1g_2,k)=\beta(g_1,k)\beta(g_2,s(g_1)^{-1}ks(g_1)).
\end{equation}
Meanwhile a cocycle $\beta$ is said to be coexact if there exists a homomorphism $\psi:K\rightarrow U(1)$ such that $\beta(g,k)=\psi(s(g)^{-1}ks(g)k^{-1})$, and the equivalence classes of cocycles modulo exact cocycles defines $H^1(G,H^1(K,U(1)))$.  The map $d_2:H^1(G,H^1(K,U(1)))\rightarrow H^3(G,U(1))$ is then given in terms of a map (also denoted $d_2$) from $Z^1(G,H^1(K,U(1)))$ to $Z^3(G,U(1))$ defined as
\begin{equation}
    d_2\beta(g_1,g_2,g_3)=\beta(g_1,s(g_1)s(g_2)s(g_3)s(g_2g_3)^{-1}s(g_1)^{-1}).
\end{equation}
Then $d_2[\beta]:=[d_2\beta]\in H^3(G,U(1))$, and it can be checked that this is well-defined as a homomorphism of cohomology groups.  Note that this is essentially the cup product of $[\beta]$ with the extension class $[\bar{c}]$, $d_2[\beta]=[\beta]\cup[\bar{c}]$, or at the level of cocycles, $d_2\beta=\beta\cup\bar{c}$.

The spectral sequence tells us that the image of $d_2$ are classes that do not pull back to nontrivial classes in $H^3(\Gamma,U(1))$.  Indeed, given $\beta\in Z^1(G,H^1(K,U(1)))$ such that $d_2\beta=\alpha$, then we can define $j\in C^2(\Gamma,U(1))$ by
\begin{equation}
\label{eq:jFromBeta}
    j(\gamma_1,\gamma_2)=\beta(\pi(\gamma_1),s(\pi(\gamma_1))\gamma_2s(\pi(\gamma_2))^{-1}s(\pi(\gamma_1))^{-1}),
\end{equation}
and one can check explicitly that $dj=\pi^\ast\alpha$.

In this case we can gauge the group $\Gamma$.  The details of the gauging will depend on the ``quantum symmetry phases''~\cite{Robbins:2021ibx} specified by $\beta$, so we denote the gauged theory as $[X/\Gamma]_{\beta}$.  This theory is well-defined, in the sense that $\Gamma$ has no
gauge anomaly, and as argued in \cite{Robbins:2021xce}, applying decomposition \cite{Hellerman:2006zs,Sharpe:2022ene}, $[X/\Gamma]_{\beta}$ is equivalent to a disjoint union of orbifolds of $X$ by nonanomalous subgroups of $G$.

Intuitively, the contribution coming from $\beta\cup\bar{c}$ cancels out the $G$ anomaly $\alpha$.  The physical interpretation is that we have taken a theory $X$ with an anomalous $G$ symmetry
and coupled it to an SPT for the $K$ symmetry, with $\beta$ describing the mixed anomaly between $G$ and $K$.  The resulting theory has an overall non-anomalous symmetry given by $\Gamma$, which can be gauged.

Of course, we are well aware that group-like symmetries are not the end of the story, and we might wonder how this picture generalizes to include non-invertible symmetries.  The simplest case is that we again begin with an anomalous group-like symmetry, but extend it in such a way that the result is a non-invertible symmetry.  The obvious issue with trying to apply the setup described above is that in this case we do not have all of the tools of group cohomology at our disposal -- mainly, we do not have a cohomological (or otherwise) classification of `extensions' that result in fusion categorical symmetries.  As a result, we'll look for an alternative description of this phenomenon that will be more amenable to generalization.

\subsection{Example: Anomalous $\Z_2$ extended to $\Z_4$}
\label{sect:ex:anomz2:z4}

As the simplest example of the above construction, let us begin with an anomalous $\Z_2$ symmetry, described by the non-trivial class $\alpha \in H^3(\Z_2,U(1))=\Z_2$.  As described in e.g.~\cite[section 4.1.1]{Robbins:2021xce}, a simple set of choices for the resolution is to take
\begin{enumerate}
    \item an extension of $G = {\mathbb Z}_2$ by $K = {\mathbb Z}_2$ to $\Gamma = {\mathbb Z}_4$,
    \item take the mixed anomaly or ``quantum symmetry phase''
    $\beta \in H^2(G,\hat{K}) = H^1({\mathbb Z}_2,{\mathbb Z}_2) = {\mathbb Z}_2$ to be the nontrivial element.
\end{enumerate}
It is straightforward to check that with these choices, $c \cup \beta = \alpha$.
Note that
non-degeneracy of the cup product requires that the extension class in $H^2(\Z_2,\Z_2)=\Z_2$ be nontrivial, which is consistent with the choice $\Gamma = \Z_4$.  There is then a non-trivial mixed anomaly valued in $H^1(\Z_2,\Z_2)=\Z_2$ between the original, anomalous $\Z_2$ symmetry and the extending, trivially-acting $\Z_2$.

\subsection{Relative phases and mixed anomalies}

As just reviewed,
in the previous papers \cite{Robbins:2021xce,Robbins_2021,Robbins:2021ibx,Robbins:2022wlr}, to implement the Wang-Wen-Witten anomaly cancellation \cite{Wang:2017loc}, in addition to enlarging the gauge group, one also turns on `relative' phases (``quantum symmetries''), analogous to discrete torsion.
In an orbifold $[X/\Gamma]$, these phases arise when a central subgroup $K \subset \Gamma$ acts trivially.
They are are classified by elements of $H^1(G,\hat{K})$, where $G = \Gamma/K$.  In the application to anomaly resolution,
they are chosen so that the image of
\begin{equation}
\label{bphases}
    d_2: \: H^1(G, \hat{K}) \longrightarrow H^3(G,U(1))
\end{equation}
is the anomaly.  

Let us review these relative phases in more detail.  The idea is that these define relative phases between contributions to the orbifold that would otherwise be identical.
For example, since $K$ acts trivially, ordinarily one has
\begin{equation}
    {\scriptstyle g} \square_h \: = \:
    {\scriptstyle gz} \square_h \: = \:
    {\scriptstyle g} \square_{zh} \: = \:
    {\scriptstyle gz} \square_{hz}
\end{equation}
for a commuting pair $g, h \in \Gamma$ and $z \in K$.  The phases $\beta$ weight the sectors in the list above,
so tthat they are no longer equal.
For example,
\begin{equation}  \label{eq:quantsymm}
{\scriptstyle g z} \square_h \: = \: \beta(\pi(h), z) \left(
{\scriptstyle g} \square_h \right),
\: \: \:
{\scriptstyle g} \square_{hz} \: = \: \beta( \pi(g), z)^{-1} \left(
{\scriptstyle g} \square_h \right),
\end{equation}
for $\pi: \Gamma \rightarrow G = \Gamma/K$ the projection, where we have used the fact that
an 
element $\beta \in H^1(G, \hat{K})$ is equivalent to a map
\begin{equation}
    \beta: \: G \times K \: \longrightarrow \: U(1).
\end{equation}

Let us describe these phases explicitly in the case of 
an anomalous $\Z_2$ extended to $\Z_4$ described in section~\ref{sect:ex:anomz2:z4}. 
Because the $K = \Z_2$ subgroup of $\Gamma = \Z_4$ is trivially-acting, we can identify $\Z_4$ partial traces with the torus partial traces of the underlying effectively-acting $\Z_2$. 
In this case, we let $\beta$ be the (only) nontrivial element of
\begin{equation}
    H^1({\mathbb Z}_4/{\mathbb Z}_2, \hat{\mathbb Z}_2) \: \cong \: {\mathbb Z}_2.
\end{equation}
Let $m, n \in \{0, \cdots, 3\}$, then we can describe these phases explicitly in
partial traces $Z_{m,n}$ as illustrated in table~\ref{table:z4z2:ex-phases},
where
\begin{equation}
    Z_{m,n} \: = \: {\scriptstyle m} \square_n .
\end{equation}

\begin{table}
    \begin{center}
        \begin{tabular}{c|c}
        ${\mathbb Z}_4$  &  ${\mathbb Z}_2$ \\ \hline
        $Z_{0,1}$ & $+ Z_{0,1}$,\\
$Z_{0,3}$ & $+ Z_{0,1}$,\\
$Z_{1,0}$ & $+ Z_{1,0}$,\\
$Z_{1,1}$ & $+ Z_{1,1}$,\\
$Z_{1,2}$ & $- Z_{1,0}$,\\
$Z_{1,3}$ & $- Z_{1,1}$,\\

$Z_{2,1}$ & $- Z_{0,1}$,\\
$Z_{2,3}$ & $- Z_{0,1}$,\\
$Z_{3,0}$ & $+ Z_{1,0}$,\\
$Z_{3,1}$ & $- Z_{1,1}$,\\
$Z_{3,2}$ & $- Z_{1,0}$,\\
$Z_{3,3}$ & $+ Z_{1,1}$.
\end{tabular}
\caption{Table of ${\mathbb Z}_4$ partial traces with trivially-acting ${\mathbb Z}_2 \subset {\mathbb Z}_4$ and nontrivial quantum symmetry.  Each ${\mathbb Z}_4$ partial trace matches a ${\mathbb Z}_4/{\mathbb Z}_2 \cong {\mathbb Z}_2$ partial trace up to a phase, determined by the quantum symmetry.
\label{table:z4z2:ex-phases}
}
    \end{center}
\end{table}

We can understand these phases at the level of diagrams decorated by topological defect lines (TDLs) by using the (non-trivial) local operators bound to the trivially-acting symmetry lines.  This procedure requires pulling the local operators through the lines for the effectively-acting symmetry, and here is where the phases from above enter, as the crossing of these operators may generate a phase. 
We will describe such alternative approaches to these `quantum symmetry' phases
in examples in sections~\ref{sect:mixedanom:symtft}, \ref{sect:z2:z4:mixedanoom}.

So far we have briefly reviewed the anomaly resolution construction for ordinary groups.  Next, we turn to constructions for more general cases.

\section{General case: SymTFT construction}   \label{sect:genl}

First, let us explain what we mean by resolving a (not-necessarily-invertible) anomalous symmetry in the language of SymTFTs.
For a given anomalous global symmetry, labeled by a fusion category $\mathcal{D}$,
its resolution corresponds to embedding the $\mathcal{D}$-symmetric theory into a larger non-anomalous symmetry as an \emph{intrinsically gapless SPT} (igSPT) phase. The notion of an igSPT phase is described in e.g.~\cite{Scaffidi:2017ppg,Thorngren:2020wet,Wen:2022tkg,Li:2023knf,Huang:2023pyk,Wen:2023otf,Bhardwaj:2024qrf} and is defined as follows.

\emph{For a fusion category $\mathcal{C}$, a $\mathcal{C}$-symmetric igSPT phase is a gapless phase protected by $\mathcal{C}$ symmetry that cannot be deformed to a gapped SPT phase.}

The defining properties of igSPT phases naturally match our anomaly resolution interest:
\begin{itemize}
    \item Gapless: The theory whose anomaly we would like to resolve is a general QFT instead of a TQFT.
    \item Intrinsically: The non-deformability to a gapped SPT implies the anomaly of the original $\mathcal{C}$-symmetry, due to one of the defining property of the anomaly that it obstructs a gapped SPT.
    \item SPT: The theory after anomaly resolution enjoys/is protected by a $\mathcal{D}$-symmetry, and does not decompose. If it decomposes, then it would be an SSB phase instead of an SPT. Again recall that the defining property of anomaly is obstruction to a SPT phase. Therefore a genuine resolution for the anomaly is to make the resolved theory have a single universe (unique ground state in the context of TQFT).
\end{itemize}
(Note that igSPT phases are only defined in condensable algebras, which are
more specialized than merely special symmetric Frobenius algebras.  So, implicit in the definition is that the larger nonanomalous symmetry is condensable,
not merely gaugeable (special symmetric Frobenius).
See Appendix~\ref{app:condensable} for more information on the distinction between condensable and gaugeable algebras.)

\subsection*{Extending the SymTFT:} 
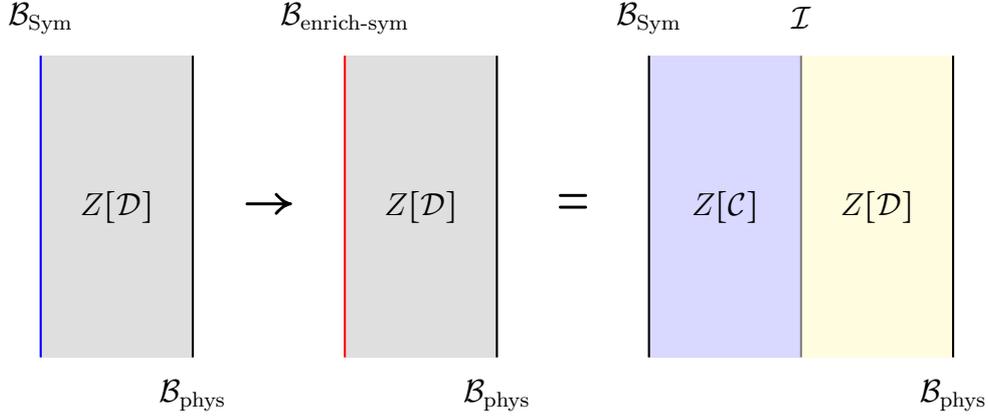
\begin{figure}[h]
    \centering
    \begin{tikzpicture}
    \fill[gray! 25] (-2,0) rectangle (0,4);
    \fill[gray! 25] (2,0) rectangle (4,4);
    \fill[blue! 15] (6,0) rectangle (8,4);
    \fill[yellow! 15] (8,0) rectangle (10,4);
    
        \draw[thick, blue] (-2,0) -- (-2,4);
        \draw[thick, black] (0,0) -- (0,4);

        \node at (1, 2) {\Huge $\rightarrow$};

        \draw[thick, red] (2,0) -- (2,4);
        \draw[thick, black] (4,0) -- (4,4);

        \node at (5,2) {\Huge $=$};

        \draw[thick, black] (6,0) -- (6,4);
        \draw[thick, gray] (8,0) -- (8,4);
        \draw[thick, black] (10,0) -- (10,4);

    \node at (-2,4.5) {$\mathcal{B}_{\text{Sym}}$};
    \node at (0,-0.5) {$\mathcal{B}_{\text{phys}}$};
    \node at (-1,2) {$Z[\mathcal{D}]$};

    \node at (2,4.5) {$\mathcal{B}_{\text{enrich-sym}}$};
    \node at (3,2) {$Z[\mathcal{D}]$};
    \node at (4,-0.5) {$\mathcal{B}_{\text{phys}}$};

    \node at (6,4.5) {$\mathcal{B}_{\text{Sym}}$};
    \node at (7,2) {$Z[\mathcal{C}]$};
    \node at (8,4.5) {$\mathcal{I}$};
    \node at (9,2) {$Z[\mathcal{D}]$};
    \node at (10,-0.5) {$\mathcal{B}_{\text{phys}}$};
    \end{tikzpicture}
    \caption{Extend the symmetry boundary of the SymTFT for an anomalous symmetry $\mathcal{C}$ to the SymTFT for a non-anomalous one $\mathcal{D}$.}
    \label{fig:symtftresolution}
\end{figure}

An illustration is shown in Figure \ref{fig:symtftresolution}. The starting point is a 2D theory with anomalous symmetry $\mathcal{D}$. In the SymTFT picture, this is a conventional sandwich construction for the Drinfeld center $Z[\mathcal{D}]$ with a physical boundary and a symmetry boundary. The anomaly resolution is realized by enriching the symmetry boundary, so that the construction is equivalent to the club sandwich configuration \cite{Bhardwaj:2023bbf}\footnote{The club sandwich can be viewed as a special case of the general `SymTree’ configuration \cite{Baume:2023kkf}, in which multiple SymTFTs are connected through (not necessarily topological) junctions, forming a tree-like structure.}, with another 3D TFT labeled by the Drinfeld center $Z[\mathcal{C}]$. The leftmost symmetry boundary $\mathcal{B}_{\text{sym}}$ is specified by the Lagrangian algebra corresponding to the non-anomalous $\mathcal{C}$ symmetry in 2D, for which we denote as $\mathcal{B}_{\text{sym}}=\mathcal{L}_\mathcal{C}$. The topological interface $\mathcal{I}$ between the two 3D TFTs is specified by a condensable algebra of $Z[\mathcal{C}]$, so that gauging/condensing this algebra, the 3D TFT $Z[\mathcal{C}]$ is reduced to the TFT $Z[\mathcal{D}]$. Given that the physical boundary is not necessarily topological, and the $\mathcal{D}$ symmetry is anomalous, the resulting 2D theory after resolution can be regarded as an $\mathcal{C}$-symmetric igSPT phase \cite{Bhardwaj:2024qrf}. 

There are some topological interface conditions defined on $\mathcal{I}$, which prevents some of topological line operators in $Z[\mathcal{C}]$ from going into the reduced theory $Z[\mathcal{D}]$. 
\begin{figure}[h]
    \centering
    \begin{tikzpicture}[scale=1.5]
        \fill[blue!20] (0,0,0) -- (2,0,0) -- (2,2,0) -- (0,2,0) -- cycle;
        \fill[blue!05] (0,0,2) -- (0,2,2) -- (2,2,2) -- (2,0,2) -- cycle;
        \fill[blue!20] (0,0,0) -- (0,0,2) -- (2,0,2) -- (2,0,0) -- cycle;
        \fill[blue!20] (0,0,0) -- (0,2,0) -- (0,2,2) -- (0,0,2) -- cycle;
        \fill[blue!20] (0,2,0) -- (2,2,0) -- (2,2,2) -- (0,2,2) -- cycle;
        \fill[blue!20] (2,0,0) -- (2,0,2) -- (2,2,2) -- (2,2,0) -- cycle;
        
        \fill[green!30] (2,0,0) -- (4,0,0) -- (4,2,0) -- (2,2,0) -- cycle;
        \fill[green!15] (2,0,2) -- (2,2,2) -- (4,2,2) -- (4,0,2) -- cycle;
        \fill[green!50] (2,0,0) -- (2,0,2) -- (4,0,2) -- (4,0,0) -- cycle;
        \fill[green!50] (2,0,0) -- (2,2,0) -- (2,2,2) -- (2,0,2) -- cycle;
        \fill[green!30] (2,2,0) -- (4,2,0) -- (4,2,2) -- (2,2,2) -- cycle;
        \fill[green!30] (4,0,0) -- (4,0,2) -- (4,2,2) -- (4,2,0) -- cycle;
        
        \draw[thick] (0,0,0) -- (2,0,0) -- (2,2,0) -- (0,2,0) -- cycle;
        \draw[thick] (0,0,0) -- (0,0,2) -- (0,2,2) -- (0,2,0) -- cycle;
        \draw[thick] (0,0,0) -- (2,0,0) -- (2,0,2) -- (0,0,2) -- cycle;
        \draw[thick] (0,2,0) -- (2,2,0) -- (2,2,2) -- (0,2,2) -- cycle;
        \draw[thick] (2,0,0) -- (2,0,2) -- (2,2,2) -- (2,2,0) -- cycle;
        
        \draw[thick] (2,0,0) -- (4,0,0) -- (4,2,0) -- (2,2,0) -- cycle;
        \draw[thick] (2,0,0) -- (2,0,2) -- (2,2,2) -- (2,2,0) -- cycle;
        \draw[thick] (2,2,0) -- (4,2,0) -- (4,2,2) -- (2,2,2) -- cycle;
        \draw[thick] (4,0,0) -- (4,0,2) -- (4,2,2) -- (4,2,0) -- cycle;
        \draw[thick] (2,0,2) -- (4,0,2) -- (4,2,2) -- (2,2,2) -- cycle;
        
        \draw[thick, black] (2,0,0) -- (2,0,2);
        \draw[thick, black] (2,2,0) -- (2,2,2);
        \draw[thick, black] (4,0,0) -- (4,0,2);
        \draw[thick, black] (4,2,0) -- (4,2,2);
        
        \node at (1, 1.7, 0) {$Z[\mathcal{C}]$};
        \node at (3, 1.7, 0) {$Z[\mathcal{D}]$};
        \node at (1, 0.6, 0) {$L_i$};
        \node at (-1.1, 0.9, 0.8) {$\mathcal{B}_{\text{sym}}=\mathcal{L}_\mathcal{C}$};
        \node at (2, 0.9, 0.8) {$\mathcal{I}$};
        \node at (4, 0.9, 0.8) {$\mathcal{B}_{\text{phys}}$};
        \draw[thick, blue] (0.5,1,0) .. controls (0.5,1,1) and (1.5,0.5,1) .. (2,0.5,1);
        
        \filldraw[blue] (2,0.5,1) circle (2pt);
    \end{tikzpicture}
    \caption{The topological line $L_i$ in $Z[\mathcal{C}]$ terminates at the topological interface $\mathcal{C}$.}
    \label{fig: line stopping at interface}
\end{figure}
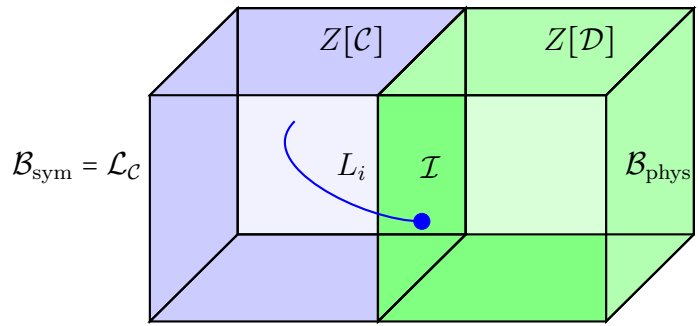
This means that for some topological line defects generating the $\mathcal{C}$ symmetry, some of their charged defects are missing in the $\mathcal{D}$ theory (see Figure \ref{fig: line stopping at interface}). This recovers the trivially acting part in the $\mathcal{C}$ symmetry via extending from $\mathcal{D}$. 
\paragraph{Decomposition after gauging the resolved symmetry:}After the anomaly resolution, one can then gauge the $\mathcal{C}$ symmetry by changing the $\mathcal{B}_{\text{sym}}=\mathcal{L}_\mathcal{C}$ into the one associated with the magnetic dual Lagrangian algebra. Denote the dual category as $\mathcal{C}'$, we can then label the new symmetry boundary as $\mathcal{B}'_{\text{sym}}=\mathcal{L}_{\mathcal{C}'}$. Under this new topological boundary condition, some line defects in $Z[\mathcal{C}]$ will terminate on both the topological boundary and the topological interface (see Figure \ref{fig: line endding on both sides}).
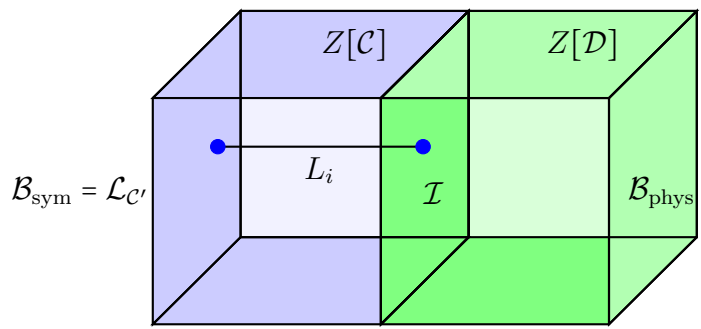
\begin{figure}[h]
    \centering
    \begin{tikzpicture}[scale=1.5]
        \fill[blue!20] (0,0,0) -- (2,0,0) -- (2,2,0) -- (0,2,0) -- cycle;
        \fill[blue!05] (0,0,2) -- (0,2,2) -- (2,2,2) -- (2,0,2) -- cycle;
        \fill[blue!20] (0,0,0) -- (0,0,2) -- (2,0,2) -- (2,0,0) -- cycle;
        \fill[blue!20] (0,0,0) -- (0,2,0) -- (0,2,2) -- (0,0,2) -- cycle;
        \fill[blue!20] (0,2,0) -- (2,2,0) -- (2,2,2) -- (0,2,2) -- cycle;
        \fill[blue!20] (2,0,0) -- (2,0,2) -- (2,2,2) -- (2,2,0) -- cycle;
        
        \fill[green!30] (2,0,0) -- (4,0,0) -- (4,2,0) -- (2,2,0) -- cycle;
        \fill[green!15] (2,0,2) -- (2,2,2) -- (4,2,2) -- (4,0,2) -- cycle;
        \fill[green!50] (2,0,0) -- (2,0,2) -- (4,0,2) -- (4,0,0) -- cycle;
        \fill[green!50] (2,0,0) -- (2,2,0) -- (2,2,2) -- (2,0,2) -- cycle;
        \fill[green!30] (2,2,0) -- (4,2,0) -- (4,2,2) -- (2,2,2) -- cycle;
        \fill[green!30] (4,0,0) -- (4,0,2) -- (4,2,2) -- (4,2,0) -- cycle;
        
        \draw[thick] (0,0,0) -- (2,0,0) -- (2,2,0) -- (0,2,0) -- cycle;
        \draw[thick] (0,0,0) -- (0,0,2) -- (0,2,2) -- (0,2,0) -- cycle;
        \draw[thick] (0,0,0) -- (2,0,0) -- (2,0,2) -- (0,0,2) -- cycle;
        \draw[thick] (0,2,0) -- (2,2,0) -- (2,2,2) -- (0,2,2) -- cycle;
        \draw[thick] (2,0,0) -- (2,0,2) -- (2,2,2) -- (2,2,0) -- cycle;
        
        \draw[thick] (2,0,0) -- (4,0,0) -- (4,2,0) -- (2,2,0) -- cycle;
        \draw[thick] (2,0,0) -- (2,0,2) -- (2,2,2) -- (2,2,0) -- cycle;
        \draw[thick] (2,2,0) -- (4,2,0) -- (4,2,2) -- (2,2,2) -- cycle;
        \draw[thick] (4,0,0) -- (4,0,2) -- (4,2,2) -- (4,2,0) -- cycle;
        \draw[thick] (2,0,2) -- (4,0,2) -- (4,2,2) -- (2,2,2) -- cycle;
        
        \draw[thick, black] (2,0,0) -- (2,0,2);
        \draw[thick, black] (2,2,0) -- (2,2,2);
        \draw[thick, black] (4,0,0) -- (4,0,2);
        \draw[thick, black] (4,2,0) -- (4,2,2);
        
        \draw[thick, black] (-0.2,0.8,0) -- (1.6,0.8,0);
        \filldraw[blue] (-0.2,0.8,0) circle (1.8pt);
        \filldraw[blue] (1.6,0.8,0) circle (1.8pt);
        
        \node at (1, 1.7, 0) {$Z[\mathcal{C}]$};
        \node at (3, 1.7, 0) {$Z[\mathcal{D}]$};
        
        \node at (-1.1, 0.7, 0.8) {$\mathcal{B}_{\text{sym}}=\mathcal{L}_{\mathcal{C}'}$};
        \node at (2, 0.7, 0.8) {$\mathcal{I}$};
        \node at (4, 0.7, 0.8) {$\mathcal{B}_{\text{phys}}$};
        \node at (1, 0.9, 0.8) {$L_i$};

    \end{tikzpicture}
    \caption{After gauging $\mathcal{C}$ symmetry to the dual symmetry $\mathcal{C}'$, some lines are terminating at both the symmetry boundary and the interface.}
    \label{fig: line endding on both sides}
\end{figure}
After shrinking the $Z[\mathcal{C}]$ slab, this topological line is localized on the symmetry boundary for $Z[\mathcal{D}]$, building a topological local operator in the resulting 2D theory (see Figure \ref{fig: top local in symtft}), 
\begin{figure}[h]
    \centering
    \begin{tikzpicture}[scale=1.5]
    \fill[green!30] (0,0) -- (2,0) -- (2,2) -- (0,2) -- cycle;
    \fill[green!15] (0,0) -- (0,2) -- (2,2) -- (2,0) -- cycle;
    \fill[green!50] (0,0) -- (0,0,-2) -- (2,0,-2) -- (2,0) -- cycle;
    \fill[green!50] (0,0) -- (0,2,0) -- (0,2,-2) -- (0,0,-2) -- cycle;
    \fill[green!30] (0,2) -- (2,2) -- (2,2,-2) -- (0,2,-2) -- cycle;
    \fill[green!30] (2,0) -- (2,0,-2) -- (2,2,-2) -- (2,2) -- cycle;
    
    \draw[thick] (0,0) -- (2,0) -- (2,2) -- (0,2) -- cycle;
    \draw[thick] (0,0) -- (0,0,-2) -- (0,2,-2) -- (0,2) -- cycle;
    \draw[thick] (0,0) -- (2,0) -- (2,0,-2) -- (0,0,-2) -- cycle;
    \draw[thick] (0,2) -- (2,2) -- (2,2,-2) -- (0,2,-2) -- cycle;
    \draw[thick] (2,0) -- (2,0,-2) -- (2,2,-2) -- (2,2) -- cycle;
    \filldraw[blue] (0.5,1.1,0.3) circle (2pt);
    
        \node at (1.2, 2.0, -0.8) {$Z[\mathcal{D}]$};
        \node at (-0.4, 0.9, 0.8) {$\mathcal{B}_{\text{enrich-sym}}$};
        \node at (0, 0.7, -1.5) {$\mathcal{O}_i$};
        \node at (3, 1.9, 1) {$\mathcal{B}_{\text{phys}}$};
\end{tikzpicture}
\caption{Shrinking the SymTFT slab for $Z[\mathcal{C}]$, a topological local operator is engineered. The resulting 2D theory thus decomposes.}
\label{fig: top local in symtft}
\end{figure}
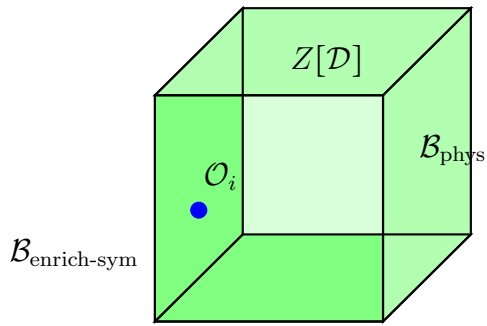
which is responsible for the decomposition under gauging trivially acting symmetries.

There is an alternative route we can take to understand the anomaly resolution using the SymTFT, which is to introduce additional couplings within the same SymTFT to engineer trivially acting symmetries in the absolute theory \cite{Lin:2025oml}. We have discussed this in great detail in section~\ref{sec:trivial_sym}, for the context of group-like symmetries.

In passing, we point out a minor subtlety.  If the two fusion categories are Morita-equivalent, their Drinfeld center are isomorphic and will correspond to the same SymTFT. Therefore, one might attempt to say, e.g., if we can resolve an anomaly by a larger group $G$, it is also true that we can resolve it by the fusion category Rep$(G)$. However, if the resulting theory with $G$ symmetry does not decompose (or has a unique ground state as a topological phase), the theory resulting from the Rep$(G)$ resolution may decompose or have multiple vacua. Due to one of the key properties of 't Hooft anomaly is obstruction to the trivially gapped phase, therefore, in this paper, we focus on the resolution where the resulting theory does not decompose (before doing any gauging). That is to say, with respect to the larger nonanomalous symmetry, the anomalous theory we start with is a igSPT phase \cite{Bhardwaj:2024qrf}.

\section{General case: Quasi-Hopf algebra formulation}   \label{sect:genl:quasihopf}

Before discussing how one resolves anomalies in an algebraic setting,
let us begin by characterizing obstructions to gauging noninvertible symmetries.
Briefly, given a noninvertible symmetry defined by some fusion category,
in order to gauge it the fusion category needs to contain a special symmetric
Frobenius algebra object.  One way to guarantee the existence of a special symmetric Frobenius algebra object is if the fusion category admits a fiber functor.
However, there exist examples of fusion categories without fiber functors that still admit special symmetric Frobenius algebra objects (for example, the Fibonacci category).

So, existence of a fiber functor is not the same thing as gaugeability.

That said, for our purposes, we will focus on cases where anomalies can be understood as a lack of a fiber functor obstructing existence of a special symmetric Frobenius algebra object, and hence obstructing gaugeability.  Then,
as 
defined in Section~\ref{sect:genl}, an anomaly resolution is an exact sequence\footnote{As explained in Appendix~\ref{app:exactseq}, the definition of an exact sequence of fusion categories we adopt in this paper is that of \cite{BN11}. However, there exists a more general definition, due to \cite{EG17}, which does not require $\cal K$ to admit a fiber functor.}  of fusion categories
\begin{equation}
    {\cal K} \: \longrightarrow \: {\cal C} \: \longrightarrow \: {\cal D},
\end{equation}
where $\cal D$ does not admit a fiber functor but $\cal C$ does.

In the study of fusion categories that do not admit a fiber functor, we often encounter two main cases. In one case, there are fusion categories whose objects have non-integer quantum dimensions. A category $\cal C$ of this kind does not even admit a \textit{linear} functor $F:{\cal C}\to \text{Vec}$, as the image $F(c)$ of any object $c\in\text{ob}({\cal C})$ needs to be a $q_c$-dimensional vector space for $q_c=\text{dim}_{\cal C}(c)$ the quantum dimension of $c$. Such is the case of the Ising category. In the other case, there are fusion categories which do admit a linear functor $F:{\cal C}\to \text{Vec}$ to the category of vector spaces, but which lack a monoidal structure, meaning a natural isomorphism $J_{X,Y}:F(X\otimes Y)\to F(X)\otimes F(Y)$ satisfying axioms that make the tensor product in $\cal C$ compatible with that in Vec. A familiar example of this is the category $\text{Vec}(G,\alpha)$ for $\alpha\in Z^3(G,U(1))$ a group 3-cocycle with nontrivial cohomology class.

\subsection{Quasi-Hopf algebras and anomalies}
Given that a vast array of fusion categories $\cal C$ equipped with a fiber functor $(F,J):{\cal C}\to \text{Vec}$ are given by representation categories of finite-dimensional semisimple Hopf algebras, one can ask what an appropriate analogue is for fusion categories admitting only a \textit{linear} functor $F:{\cal C}\to \text{Vec}$ (not necessarily with a monoidal structure $J$). In this case, examples are now given by representation categories of \textit{quasi}-Hopf algebras.

The definition of a quasi-Hopf algebra $H$ is similar to that of a Hopf algebra, with the key difference that the comultiplication is now coassociative \textit{up to} a nontrivial element $\Phi\in H^{\otimes 3}$, the \textit{coassociator}. As we will explain below, the coassociator is precisely what encodes the obstruction to the existence of a monoidal structure $J$ on the forgetful functor
\begin{equation}
    F:  \: \text{Rep}(H) \: \longrightarrow \: \text{Vec}
\end{equation}
from the tensor category of representations of $H$ to the category of vector spaces. Intuitively, the 't Hooft anomaly of $\text{Rep}(H)$ is classified by the quasi-Hopf algebra coassociator $\Phi$, in the same sense that the 't Hooft anomaly of $\text{Vec}(G,\alpha)$ is classified by the 3-cocycle $\alpha$.

Of course, a category $\text{Vec}(G,\alpha)$ is non-anomalous not only if $\alpha$ is strictly trivial but more generally if its cohomology class is trivial. In the case of quasi-Hopf algebras, this equivalence relation is provided by the notion of \textit{gauge equivalence} of coassociators, recalled below. If the coassociator $\Phi$ of a quasi-Hopf algebra $H$ is gauge equivalent to the trivial coassociator $\Phi'=1$, then $\text{Rep}(H)$ admits a fiber functor. Simply put, the gauge twist turns the quasi-Hopf algebra $H$ into a Hopf algebra. In this case, one usually says $H$ is gauge equivalent to a Hopf algebra.

In summary, a quasi-Hopf algebra $H$ with a coassociator $\Phi$ describes a fusion category $\text{Rep}(H)$ whose 't Hooft anomaly is classified by $[\Phi]_{\rm gauge}$ the gauge equivalence class of $\Phi$.

\subsection{Exact sequences and anomaly resolutions}

Now, recall that one way to obtain exact sequences of non-anomalous fusion categories is by working with exact sequences of Hopf algebras
\begin{equation}
    A \: \longrightarrow \: \tilde{H} \: \longrightarrow \: H
\end{equation}
and applying the representation functor $\text{Rep}(-)$ to get an exact sequence of fusion categories
\begin{equation}
    \text{Rep}(H) \: \longrightarrow \: \text{Rep}(\tilde{H}) \: \longrightarrow \: \text{Rep}(A).
\end{equation}
In light of the intuition of quasi-Hopf algebras and the gauge equivalence relation on coassociators, whose definitions we recall below, we can extend this method to obtain anomaly resolutions. Namely, given an exact sequence of quasi-Hopf algebras
\begin{equation}
    D \: \hookrightarrow \: C \: \longrightarrow \: K
\end{equation}
where $D$ is \textit{not} gauge equivalent to a Hopf algebra but $C$ and $K$ are, one applies the representation functor to get
\begin{equation}
    \text{Rep}(K) \: \longrightarrow \: \text{Rep}(C) \: \longrightarrow \: \text{Rep}(D),
\end{equation}
so that $\text{Rep}(C)$ admits a fiber functor, but not $\text{Rep}(K)$. This resolves the anomalous $\text{Rep}(K)$ by the non-anomalous $\text{Rep}(C)$.

In particular, this picture subsumes the anomaly resolution of group-like symmetries. This is because given a finite group $G$ and an anomaly 3-cocycle $\alpha$, one can construct a quasi-Hopf algebra $\C^G_{\alpha}$ which is gauge equivalent to $\C^G$ iff $\alpha$ has a trivial cohomology class. Then, a group anomaly resolution
\begin{equation}
    K \: \longrightarrow \: \tilde{G} \: \stackrel{\pi}{\longrightarrow} \:  G
\end{equation}
where the 3-cocycle $\alpha\in Z^3(G,U(1))$ with nontrivial cohomology class pulls back to a 3-coboundary $\pi^*\alpha\in Z^3(\tilde{G},U(1))$ is equivalently described by a sequence of quasi Hopf algebras
\begin{equation}
    \C^G_{\alpha} \: \longrightarrow \: (\C^{\tilde{G}})_{\rm gauge} \: \longrightarrow \: (\C^K)_{\rm gauge},
\end{equation}
where $(\C^{\tilde{G}})_{\rm gauge}, (\C^K)_{\rm gauge}$ are quasi-Hopf algebras gauge equivalent to the dual group algebras $\C^{\tilde{G}}, \C^K$. At the level of representation categories, this is simply
\begin{equation}
    \text{Vec}(K) \: \longrightarrow \: \text{Vec}(\tilde{G}) \: \longrightarrow \: \text{Vec}(G,\alpha).
\end{equation}

\subsection{Mathematical definition}
Before proceeding with the definitions, we introduce some notation, as we will be dealing with elements of tensor products of coalgebras frequently. Let $H$ be a coalgebra.  An element $a \in H^{\otimes n}$ in the $n$th tensor product of $H$ is written as a sum
\begin{equation}
    a = \sum_j a_j^{(1)}\otimes a_j^{(2)}\otimes \cdots \otimes a_j^{(n)}\in H^{\otimes n}
\end{equation}
of elements $a_j^{(1)},a_j^{(2)},\cdots, a_j^{(n)}\in H$, and $j$ is an index that labels all elements in this expansion. 

\

We now define quasi-Hopf algebras. One first defines a \textit{quasi}-bialgebra, which consists of a tuple $(H,\Delta,\epsilon,\Phi)$, where $H$ is an associative algebra, $\Delta:H\to H\otimes H$ and $\epsilon:H\to\C$ are algebra maps, called the comultiplication, and counit, respectively, and $\Phi\in H\otimes H\otimes H$ is an invertible element in the tensor algebra, called the \textit{coassociator}. The coassociator and its inverse are denoted as  $\Phi=\sum_j X_j^{(1)}\otimes X_j^{(2)}\otimes X_j^{(3)}$ and $\Phi^{-1}=\sum_j x_j^{(1)}\otimes x_j^{(2)}\otimes x_j^{(3)}$ (with the sum over elements assumed).

The tuple $(H,\Delta,\epsilon,\Phi)$ satisfies the following identities:
\begin{enumerate}
    \item counit \begin{equation}\label{eq:qcounit}
    (\text{Id}_H\otimes\epsilon)(\Delta(h)) = (\epsilon\otimes\text{Id}_H)(\Delta(h))=h,
\end{equation}
    \item quasi-coassociativity
    \begin{equation}\label{eq:qcoass}
    (\text{Id}_H\otimes\Delta)(\Delta(h)) = \Phi (\Delta\otimes\text{Id}_H)(\Delta(h))\Phi^{-1},
\end{equation}
    \item normalized 3-cocycle conditions
    \begin{equation}\label{eq:q3cocycle2}
    (\text{Id}_H\otimes\epsilon\otimes\text{Id}_H)\Phi =1_H\otimes 1_H,
\end{equation}
\begin{equation}\label{eq:q3cocycle1}
    (1_H\otimes\Phi)(\text{Id}_H\otimes \Delta \otimes \text{Id}_H)(\Phi)(\Phi\otimes 1_H)=(\text{Id}_H\otimes \text{Id}_H\otimes \Delta)(\Phi)(\Delta\otimes \text{Id}_H\otimes\text{Id}_H)(\Phi).
\end{equation}
\end{enumerate}

A quasi-Hopf algebra is a quasi-bialgebra $(H,\Delta,\epsilon,\Phi)$ endowed with an algebra anti-homomorphism $S:H\to H$ (the antipode), and a pair of distinguished elements $\alpha,\beta\in H$ such that $\epsilon(\alpha)=\epsilon(\beta)=1$, satisfying for all $h\in H$,
\begin{eqnarray}
    \sum_{j} S((\Delta(h))_j^{(1)})\,\alpha\,(\Delta(h))_j^{(2)} = \epsilon(h)\alpha,\\
 \sum_j (\Delta(h))_j^{(1)}\beta  S((\Delta(h))_j^{(2)}) = \epsilon(h)\beta,\end{eqnarray}
\begin{eqnarray}
    \sum_j X_j^{(1)}\beta S(X_j^{(2)}) \alpha X_j^{(3)} = 1_H, \\ 
 \sum_j S(x_j^{(1)})\alpha x_j^{(2)} \beta S(x_j^{(3)}) = 1_H.
\end{eqnarray}

Clearly, Hopf algebras are special cases of quasi-Hopf algebras, where $\Phi=1\otimes 1\otimes 1$, and $\alpha=\beta=1$.

Less trivial examples of quasi-Hopf algebras are given by pairs $(G,\alpha)$ for $G$ a finite group and $\alpha\in Z^3(G,U(1))$ a group 3-cocycle. The bialgebra structure is given by the dual group algebra $\C^G$ and coassociator is simply the 3-cocycle $\omega\in \C^G\otimes \C^G\otimes \C^G$ viewed as a three-parameter $\C^{\times}$-valued function on $G$. We will denote this quasi-Hopf algebra as $\C^G_{\alpha}$.

Now, we come to the notion of gauge twist, which as we argued above, is the correct equivalence relation for classifying 't Hooft anomalies in representation categories of quasi-Hopf algebras. Given a quasi-Hopf algebra $(H,\Delta,\epsilon,S,\alpha,\beta)$, a \textit{gauge twist} is an invertible element $F=\sum_j F_j^{(1)}\otimes F_j^{(2)}\in H\otimes H$ with inverse $F^{-1}=\sum_j G_j^{(1)}\otimes G_j^{(2)}\in H\otimes H$ such that $\sum_j \epsilon(F_j^{(1)})F_j^{(2)}=\sum_k F_k^{(1)}\epsilon(F_k^{(2)})=1_H$. A gauge twist defines a new quasi-Hopf algebra $(H,\Delta^F,\epsilon,S,\alpha^F,\beta^F)$ with coalgebra structure
\begin{eqnarray}
    \Delta^F(h) &=& F\Delta(h) F^{-1} ,
    \\
    \Phi^F &=& (1_H\otimes F)(\text{Id}_H\otimes\Delta)(F)\Phi (\Delta\otimes\text{Id}_H)(F^{-1})(F^{-1}\otimes1_H),
\end{eqnarray}
and the distinguished elements
\begin{eqnarray}
    \alpha^F = \sum_j S(G_j^{(1)})\alpha G_j^{(2)}, \\ \beta^F = \sum_j F_j^{(1)}\beta S(F_j^{(2)}).
\end{eqnarray}

Note that the notion of a Drinfeld twist \cite{majid} is a special case of a gauge twist, with the crucial difference that while the Drinfeld twist of a Hopf algebra is again a Hopf algebra, the gauge twist of a Hopf algebra is generally only a quasi-Hopf algebra. 

Gauge twists are important for the theory of representation categories because if a pair $H_1,H_2$ of quasi-Hopf algebras are \textit{gauge equivalent}, meaning there is a gauge twist $F\in H_1\otimes H_1$ such that $H_1^F\cong H_2$, then their representation categories are \textit{tensor} equivalent $\text{Rep}(H_1)\cong \text{Rep}(H_2)$. In particular, this says that a quasi-Hopf algebra with a coassociator that is not of the form $1\otimes 1\otimes 1$ (as that of a Hopf algebra) might still have a representation category that admits a fiber functor. Intuitively, a coassociator $\Phi$ coming from a gauge twist $F$ is trivial in the same sense that a group anomaly 3-cocycle is trivial if it is a 3-coboundary.

Therefore, an anomaly resolution in terms of quasi-Hopf algebras can be stated more precisely as an exact sequence of quasi-Hopf algebras
\begin{equation}
    D\xrightarrow{\imath} C\to K,
\end{equation}
where
\begin{enumerate}
    \item $D$ is a quasi-Hopf algebra \textit{not} gauge equivalent to a Hopf algebra,
    \item $C,K$ are quasi-Hopf algebras which \textit{are} gauge equivalent to Hopf algebras.
\end{enumerate}
We can understand the anomaly resolution as the fact that the \textit{inclusion} $\imath(\Phi_D)\in C\otimes C\otimes C$ of the coassociator $\Phi_D$ of $D$ becomes \textit{trivialized} (in the sense of gauge equivalence) in $C$, since $\Phi_C\in C\otimes C\otimes C$ is gauge-equivalent to a trivial coassociator. As mentioned previously, this extends group anomaly resolutions
\begin{equation}
    K\to \tilde{G}\xrightarrow{\pi} G,
\end{equation}
where one talks about the \textit{pullback} $\pi^*\alpha$ of the anomaly 3-cocycle $\alpha \in Z^3(G,U(1))$ being cohomologically trivial in $Z^3(\tilde{G},U(1))$. We will explicitly apply these definitions in the construction in Section~\ref{sssec:algebraic}, which will allow us to deduce very general resolutions (c.f. Eq.~(\ref{eq:resolutionbyz2center})).

\section{Relation between the SymTFT and Hopf algebras}  \label{sect:reln}

In this section we discuss the relationship between the SymTFT and the algebraic
approaches to anomaly resolution, and to noninvertible symmetries more generally.

First, let us consider the SymTFT associated to a single fusion category, before discussing anomaly resolution in this language.
To begin, suppose we have a two-dimensional theory with a global $G$ symmetry, for $G$ a finite group.  The three-dimensional SymTFT bulk is a gauged $G$ theory, specifically, a three-dimensional $G$ Dijkgraaf-Witten theory.
The 't Hooft anomaly of the two-dimensional $G$ symmetry, $\alpha \in H^3(G,U(1))$, is the level of the three-dimensional Dijkgraaf-Witten theory.  

Now, suppose that 
instead of an ordinary finite group $G$ we have a fusion category ${\cal C}$.
The existence of an anomaly is reflected in the lack of a fiber functor on ${\cal C}$.
(In particular, note that the choice of fusion category implicitly encodes the anomaly, analogously to the difference between Vec$(G)$ and Vec($G, \alpha$), both of which are associated to the same finite group $G$, but which define different fusion categories.)
It is believed that the three-dimensional Dijkgraaf-Witten theory for $G$ is replaced by a Turaev-Viro TFT on the fusion category ${\cal C}$ \cite{Turaev:1992hq,Turaev:1994xb,Barrett:1993zf,Barrett:1993ab,Kirillov:2010nh,b2,b3} or equivalently a Reshetikhin-Turaev TFT on the Drinfeld center of the fusion category ${\cal C}$, see for example
\cite{Kirillov:2010nh,b2,b3}. (In the special case that ${\cal C} = {\rm Vec}(G,\alpha)$ for a finite-group $G$, the Turaev-Viro theory on ${\cal C}$ reduces to Dijkgraaf-Witten theory for the group
$G$ with twist $\alpha$.)
The idea is that the Drinfeld center contains information about topological line operators and their braiding and linking relations in three dimensions, and hence can define a TFT, as discussed in e.g.~\cite{Kirillov:2010nh,b2,b3}.
The choice of fusion category, specifically the associator in the fusion category, encodes the anomaly (analogously to the difference between Vec$(G)$ and Vec($G, \alpha$), both of which are associated to the same symmetry group $G$, but which define different fusion categories).

Next, let us turn to anomaly resolution.  In the language of SymTFT's, to resolve an anomalous fusion category ${\cal C}$, we construct a `club sandwich' \cite{Bhardwaj:2023bbf},
in which one half of the club sandwich is a SymTFT corresponding to the original fusion category ${\cal C}$, and the other half is the SymTFT corresponding to the resolution $\widetilde{\cal C}$.  Between the two halves is an interface.  In the corresponding algebraic construction,
we write 
\begin{equation}
    {\cal C} \: = \: {\rm Rep}({\cal H}), \: \: \:
    \widetilde{\cal C} \: = \: {\rm Rep}( \widetilde{\cal H}),
\end{equation}
where ${\cal H}$, $\widetilde{\cal H}$ are (quasi-)Hopf algebras.  The two halves of the club sandwich correspond to each of the terms in the sequence
\begin{equation}
   {\rm Rep} (\widetilde{\cal H}) \: \stackrel{*}{\longrightarrow} \: 
   {\rm Rep}({\cal H}),
\end{equation}
and the interface between the two halves of the club sandwich corresponds to the
morphism $*$.

In the case of ordinary groups $\tilde{G} \rightarrow G$, with kernel $K$, the interface of the
club sandwich can be understood in terms of a boundary defined by gauging the
one-form symmetry $BK = K^{(1)}$, as gauging the $\tilde{G}$ Dijkgraaf-Witten theory by $BK$ results in the $G$ Dijkgraaf-Witten theory.  For fusion categories, we expect an analogous interpretation, involving a gauged $BK$ for $K$ the kernel of the map $\widetilde{\cal C} \rightarrow {\cal C}$.

One possible formulation for the $BK$ analog for fusion categories is the suspension construction $\Sigma\mathcal{C}$ for a braided fusion category $\cal C$ presented in \cite{DR18}. Given a braided fusion category $\cal C$, one can construct a pointed presemisimple 2-category $B{\cal C}$ with a single object $\bullet$ such that $\text{End}(\bullet)={\cal C}$. The braiding of $\cal C$ allows to endow $B{\cal C}$ with a \textit{pre}fusion structure, which upon linear completion becomes a fusion 2-category $\Sigma{\cal C}$. This fusion 2-category can be equivalently formulated as the fusion 2-category of $\cal C$-module linear categories. Given a finite abelian group $K$, the fusion category ${\cal C}=\text{Vec}(K)$ is not only braided but strict symmetric, and the suspension 2-category $\Sigma\text{Vec}(K)= 2\text{Vec}(BK)$ is equivalent to the fusion 2-category of $BK$-graded 2-vector spaces.

\section{Example: Revisiting ordinary groups}   \label{sect:redo}

Previously in section~\ref{sect:rev} we reviewed anomaly resolution for ordinary groups.  In this section we will discuss how those results for ordinary groups can be understood in the language of SymTFTs and (quasi-)Hopf algebras.

\subsection{General picture in SymTFTs}

We begin by outlining how anomaly resolution for ordinary groups is described in terms of SymTFTs.  In that language, the anomaly resolution is accomplished by a club sandwich construction with one three-dimensional region contains a Dijkgraaf-Witten theory for
$\Gamma$ (at vanishing level), bordered by another three-dimensional region containing a Dijkgraaf-Witten theory for $G$ at nonzero level (anomaly).  The choice of quantum symmetry $\beta \in H^1(G,H^1(K,U(1)))$ is implicitly encoded in the topological interface between those two three-dimensional regions.

In the rest of the section, we will present a detailed case study for resolving anomalous $\mathbb{Z}_2$ by extending to $\mathbb{Z}_4$.

\subsection{Anomalous $\Z_2$ extended to $\Z_4$ via SymTFTs}
\label{sect:ex:anomz2:z4:redo}

In this subsection, we discuss how to realize the anomaly resolution for anomalous $\mathbb{Z}_2$ extended to $\mathbb{Z}_4$ from 3D SymTFTs. This amounts to extending the SymTFT for the anomalous $\mathbb{Z}_2$ into two SymTFTs slabs, one for anomalous $\mathbb{Z}_2$ and the other for $\mathbb{Z}_4$, separated by topological interface. As investigated in e.g.~\cite{Bhardwaj:2024qrf}, this bulk construction classifies the possible phases under $\mathbb{Z}_4$ symmetry, which we list below 
\begin{itemize}
	\item The $\Z_4$ SPT phase, describing a totally trivially-acting $\Z_4$ symmetry.
	\item The `canonical' $\Z_4$ gapless SPT phase, describing a fully effectively-acting $\Z_4$ symmetry.
	\item Another gapless SPT phase, describing a $\Z_4$ symmetry with trivially-acting $\Z_2$ subgroup.
	\item Finally, there is an `intrinsically gapless' SPT (igSPT) phase, which describes our situation of interest -- a $\Z_4$ symmetry with trivially-acting $\Z_2$ subgroup in which the $\Z_4/\Z_2 = \Z_2$ quotient is anomalous.
\end{itemize}
Therefore, from a $\mathbb{Z}_4$ symmetry perspective, the 2D QFT with anomalous $\mathbb{Z}_2$ symmetry is an igSPT phase.

Before moving to an explicit 3D bulk investigation, let us first briefly discuss from a purely 2D perspective for this anomaly resolution by turning on background fields\footnote{We thank Yunqin Zheng for valuable discussions on this point}. For a 2D QFT equipped with a $\mathbb{Z}_2$ symmetry, one can turn on a background field $A_1$ for this $\mathbb{Z}_2$ symmetry. If this symmetry is anomalous, then the partition function of the 2D QFT is not invariant under the $\mathbb{Z}_2$ symmetry transformation. Instead, it picks up a phase
\begin{equation}
    Z[A_1]\rightarrow Z[A_1+\delta \alpha_0]=Z[A_1]\exp \left( i\pi \int_{M_2}\alpha_0\frac{\delta A_1}{2} \right)
\end{equation}

Resolving the above anomaly amounts to curing the phase ambiguity of the partition function when coupling to the symmetry background field. To achieve that, let us first introduce another $\mathbb{Z}_2$ background field $B_1$, and stack a $\mathbb{Z}_2\times \mathbb{Z}_2$ SPT phase onto the 2D QFT. The resulting partition function coupled to the $\mathbb{Z}_2\times \mathbb{Z}_2$ background reads
\begin{equation}\label{eq: resolved partition function for z2z4}
    Z[A_1]\exp \left( i\pi\int_{M_2}A_1\cup B_1 \right),
\end{equation}
where the second factor corresponds to a $\mathbb{Z}_2\times \mathbb{Z}_2$ SPT, or equivalently, a discrete torsion.

Let us again perform the $\mathbb{Z}_2$ symmetry transformation for $A_1$
\begin{eqnarray}
    \lefteqn{
    Z[A_1]\exp \left( i\pi\int_{M_2}A_1\cup B_1 \right)
    } \nonumber
    \\
    & \longrightarrow & 
    Z[A_1]\exp \left(  i\pi \int_{M_2}\alpha_0\frac{\delta A_1}{2}  \right) \exp \left( i\pi\int_{M_2}A_1\cup B_1 \right) \exp \left( i\pi\int_{M_2}\delta \alpha_0 \cup B_1 \right) ,
    \nonumber
    \\
    & \longrightarrow &
    Z[A_1]\exp \left( i\pi\int_{M_2}A_1\cup B_1 \right) \exp \left( i\pi  \int_{M_2}\alpha_0 ( \frac{\delta A_1}{2}-\delta B_1) \right).
    \label{eq: z2 anomaly cancelation bkg field steps}
\end{eqnarray}

Imposing the condition 
\begin{equation}\label{eq: extension from z2 to z4}
    \delta B_1=\frac{\delta A_1}{2},
\end{equation}
the partition function (\ref{eq: resolved partition function for z2z4}) will be free of phase ambiguity, and the anomaly is thus resolved. However, the condition we impose implies that the two $\mathbb{Z}_2$'s do not build a direct product, but a group extension 
\begin{equation}
    1 \: \longrightarrow \: \mathbb{Z}_2^{(B)} \: \longrightarrow \: \Z_4 \: \longrightarrow \: \Z_2^{(A)} \: \longrightarrow \: 1.
\end{equation}
The $\Z_2^{(B)}$ is trivially-acting in the sense that its background field is only involved in the discrete torsion $\exp \left( i\pi\int_{M_2}A_1\cup B_1 \right)$, thus does not act on any genuine local operators of the 2D QFT.

\subsubsection{SymTFT in terms of gauge fields}

Let us discuss how to implement the anomaly resolution above via SymTFTs. The starting point is the SymTFT for an anomalous $\mathbb{Z}_2$ symmetry, which is a twisted $\mathbb{Z}_2$ Dijkgraaf-Witten theory
\begin{equation}
    S_{\mathbb{Z}_2}^{\omega} \: = \: \int_{M_3} \left( \frac{2\pi}{2}a_1\cup \delta \hat{a}_1+\frac{2\pi}{4}a_1\cup \delta a_1 \right).
\end{equation}
As in the conventional SymTFT setup, we introduce two boundaries for this 3D bulk theory, one physical and the other topological. The physical boundary defines local information of our interested 2D QFT, while the topological boundary imposes the following boundary condition
\begin{equation}
    a_1|_{\partial} \: = \: A_1.
\end{equation}
This boundary condition specifies a $\Z_2$ global symmetry for the 2D QFT. The background field is given by the boundary condition field profile $A_1$, and the anomaly is captured by a 3D anomaly inflow $\exp \left[ i\pi \int_{M_3}A_1\cup \delta A_1 \right]$.

Extending this anomalous $\Z_2$ into a non-anomalous $\Z_4$ symmetry requires extend the 3D bulk by promoting the topological boundary for $S_{\Z_2}^\omega$ into a topological interface between $S_{\Z_2}^\omega$ and the SymTFT for $\Z_4$. In order to make the underlying group extension manifest, we express the $\Z_4$ SymTFT as a twisted $\Z_2^2$ Dijkgraaf-Witten theory
\begin{equation}\label{eq: Z_4 SymTFT}
    S_{\Z_4} \: = \: \int_{M_3} \left( \frac{2\pi}{2}a_1\cup \delta \hat{a}_1+\frac{2\pi}{2}b_1\cup \delta \hat{b}_1-\frac{2\pi}{4}a_1\cup \delta \hat{b}_1 \right).
\end{equation}
In addition to the topological interface, we introduce a topological boundary for this SymTFT. The $\Z_4$ symmetry in 2D is realized by the following boundary condition
\begin{equation}\label{eq: z4 symmetry boundary condition}
    a_1|_{\partial} \: = \:A_1,  
    \: \: \:
    b_1|_{\partial} \: = \: B_1.
\end{equation}
Naively, this gives rise to a $\Z_2^2$ symmetry. However, notice the equation of motion for $\hat{b}_1$ reads
\begin{equation}
    \delta b_1 \: = \: \frac{1}{2}\delta a_1,
\end{equation}
which reduces to equation (\ref{eq: extension from z2 to z4}) under the topological boundary condition. This implies a group extension structure for the two $\Z_2$'s and, as a result, specifies a $\Z_4$ symmetry.

The $S_{\Z_4}$ and $S_{\Z_2}^\omega$ theories are connected via gauging/condensing topological line operators. Start with $S_{\mathbb{Z}_4}$ theory, one can gauge a certain $\mathbb{Z}_2^{(1)}$ 1-form symmetry to obtain $S_{\mathbb{Z}_2^\omega}$ theory. Thus, one can perform a half-space gauging for $S_{\mathbb{Z}_4}$ theory, realizing a topological interface between $S_{\mathbb{Z}_4}$ and $S_{\mathbb{Z}_2^\omega}$. Precisely, the 1-form symmetry to be gauged is generated by the line operator $\exp (\pi i\oint a_1+\hat{b}_1 )$. Coupling the background field $c_2$ for this 1-form symmetry to the theory $S_{\mathbb{Z}_4}$ and summing it over, we derive
\begin{equation}
    \sum_{c_2}\exp(iS_{\mathbb{Z}_4})\exp \left(\pi i \int_{M_3}c_2(a_1+\hat{b}_1) \right).
\end{equation}
It is easy to see $c_2$ serves as a Lagrangian multipler, imposing the condition
\begin{equation}
    a_1+\hat{b}_1=0.
\end{equation}
Substituting the above condition into $S_{\mathbb{Z}_4}$ we have
\begin{equation}
    \exp \left( \frac{2\pi i}{2}\int_{M_3}a_1\cup \delta \hat{a}_1+\frac{1}{2}a_1\cup \delta a_1-b_1\cup \delta a_1 \right).
\end{equation}
We can then integrate by part and write down
\begin{equation}
     \exp \left( \frac{2\pi i}{2}\int_{M_3}a_1\cup \delta \hat{a}_1+\frac{1}{2}a_1\cup \delta a_1-a_1\cup \delta b_1+\frac{2\pi i}{2}\int_{M_3}\delta ({a_1\cup b_1})\right).
\end{equation}
Redefining $\hat{a}-b\rightarrow \hat{a}$, we obtain $S_{\mathbb{Z}_2}^\omega$ theory
\begin{equation}
    \exp(\frac{2\pi i}{2}\int_{M_3}a_1\cup \delta \hat{a}_1+\frac{1}{2}a_1\cup \delta a_1 )
\end{equation}
up to a 2D boundary term
\begin{equation}
    \exp \left(\pi i\int_{M_2}a_1\cup b_1 \right).
\end{equation}

\begin{figure}[H]
    \centering
    \begin{tikzpicture}[scale=1]
    
        \fill[blue! 20] (0,0) rectangle (2,6);
        \fill[yellow! 20] (2,0) rectangle (4,6); 

        \draw[thick, black] (0,0) -- (0,6);
        \draw[thick, black] (2,0) -- (2,6);
        \draw[thick, black] (4,0) -- (4,6);

        \node at (1,3) {$S_{\Z_4}$};
        \node at (3,3) {$S^\omega_{\Z_2}$};
        \node at (2,6.5) {$\mathcal{I}$};
        \node at (0,6.5) {$\mathcal{B}_{sym}$};
        \node at (4, 6.5) {$\mathcal{B}_{Phys}$};
        \node at (-0.7, 3.7) {$a_1 = A_1$}; 
        \node at (-0.7, 3.0) {$b_1 = B_1$};

        \node at (6.1,4) {$\longrightarrow$};

        \fill[yellow! 20] (8.5, 0) rectangle (10.5,6);
        \draw[thick, black] (8.5,0) -- (8.5,6);
        \draw[thick, black] (10.5,0) -- (10.5,6);

        \node at (10.5,6.5) {$\mathcal{B}_{phys}$};
        \node at (8.5,6.5) {$\mathcal{B}'_{sym}$};
        \node at (9.5, 3) {$S^\omega_{\Z_2}$};
        \node at (7.2,3) {$e^{ \left( i\pi \int_{M_2} A_1 \cup B_1 \right) } $}; 
    \end{tikzpicture}
    \caption{Shrinking the $S_{\mathbb{Z}_4}$ SymTFT slab one ends up with a nontrivial $\mathbb{Z}_2^2$ SPT stacking on the symmetry boundary of the $S_{\mathbb{Z}_2}^\omega$ theory.}
    \label{fig:z4clubsandwich}
\end{figure}
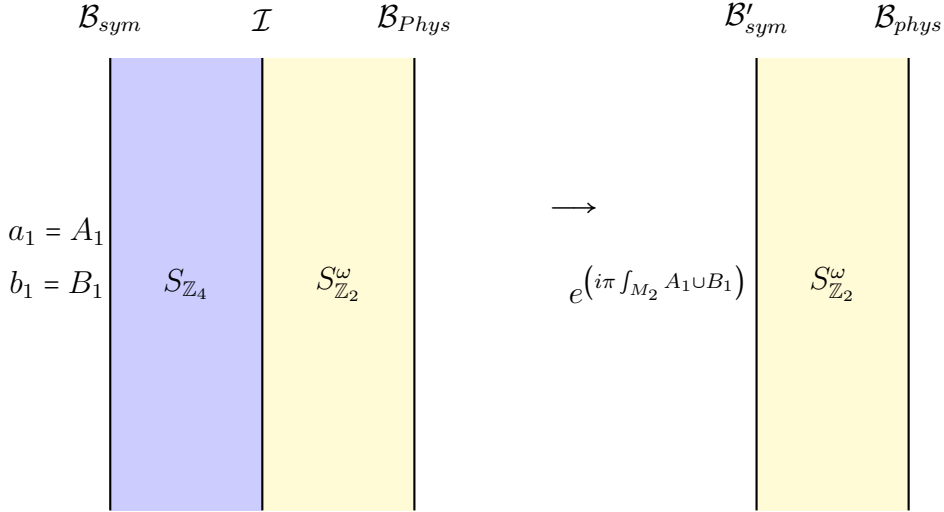

After shrinking the 3D bulk supporting the $S_{\mathbb{Z}_4}$, the above 2D boundary term subject to the boundary condition (\ref{eq: z4 symmetry boundary condition}) becomes the SPT 
\begin{equation}
    \exp \left( i\pi\int_{M_2}A_1\cup B_1 \right)
\end{equation}
stacked on the topological boundary of the  $S_{\mathbb{Z}_2}^\omega$  theory. This reproduces the discussion at the beginning of Section~\ref{sect:ex:anomz2:z4:redo}. See Figure~\ref{fig:z4clubsandwich} for an illustration. 

\subsubsection{Mixed anomaly phases from SymTFT anyons}  \label{sect:mixedanom:symtft}

Let us describe the `quantum symmetry' phases from the 3d TFT perspective. Instead of the Lagrangian description, from now on, we will use the abstract language of anyons. Since our overall (enlarged) symmetry is a non-anomalous $\Z_4$, the theory in question will have a $\mcZ(\text{Vec}_{\Z_4})$ SymTFT.  There are 16 anyons, generated by the bosonic lines $e^4=m^4=1$. That is, we have both a magnetic and electric $\Z_4$ symmetry, such that $\mcZ(\text{Vec}_{\Z_4})\simeq\Z_4\boxtimes\hat{\Z}_4$.  In the Lagrangian description of the SymTFT (\ref{eq: Z_4 SymTFT}), anyons correspond to 
\begin{equation}
\begin{split}
    e=e^{i\pi \oint b}, ~e^2=e^{i\pi \oint a}, ~e^3=e^{i\pi \oint a+b}, 
    \\ 
    m=e^{i\pi \int \hat{a}}, ~m^2=e^{i\pi \oint \hat{b}}, ~m^3=e^{i\pi \oint \hat{a}+\hat{b}}.
\end{split}
\end{equation}

 The SymTFT has both a physical and symmetry boundary.  For the symmetry boundary we choose a topological boundary condition, which in this case corresponds to a maximal (Lagrangian) algebra.  Here we will mimic the selection in \cite[Figure 1]{Bhardwaj:2024qrf}, choosing the symmetry boundary's algebra to be $\mathcal{B}_{\text{sym.}}=1\oplus e\oplus e^2\oplus e^3$, corresponding to the boundary condition (\ref{eq: z4 symmetry boundary condition}).  This means that the $e$ lines are condensed along this boundary, and can end on it.  The $m$ lines, on the other hand, furnish the symmetry boundary with its $\Z_4$ symmetry.

Upon interval compactification, some or all of this $\Z_4$ may be trivially-acting in the resulting theory.  In order for this to be the case, there must be anyons stretching from the symmetry boundary to the physical boundary which describe topological point operators (TPOs) living on the symmetry lines, allowing us to change between symmetry actions.  For instance, if we take the trivial boundary condition $1$ for the physical boundary, no anyons are condensed along that boundary and there can be no TPOs on the $\Z_4$ symmetry boundary's lines, hence the entire symmetry must act effectively.  This corresponds to the `canonical gSPT' phase of \cite[Figure 1]{Bhardwaj:2024qrf}, which describes an effectively-acting symmetry.

Another choice we could make would be to choose a topological boundary condition for the physical boundary, in which case there is the possibility of TPOs living on the symmetry boundary.  When we pick the physical boundary's condensable algebra to match the symmetry boundary (i.e.~$\mathcal{B}_{\text{phys.}}=1\oplus e\oplus e^2\oplus e^3$), the anyons attached to these TPOs end on physical boundary TPOs, leading to ground state degeneracy.  This leads to a symmetry-broken phase, which in this case is $\Z_4$ gauge theory.  The opposite extreme has us taking $\mathcal{B}_{\text{phys.}}=1\oplus m\oplus m^2\oplus m^3$, in which case the anyons connecting to the symmetry boundary's TPOs also run along the physical boundary, leading to a totally trivially-acting symmetry after interval compactification -- this produces the $\Z_4$ SPT phase.

The case of interest to us is an in between one where we take a non-trivial but also non-topological boundary condition for the physical boundary, allowing for a $\Z_4$ symmetry which is only partially effectively-acting.  The two cases in which we are able to do this without introducing ground state degeneracy are $\mathcal{B}_{\text{phys.}}=1\oplus m^2$, leading to a gSPT phase and $\mathcal{B}_{\text{phys.}}=1\oplus e^2m^2$, leading to the igSPT phase.  As discussed above, both of these lead to phases in which the effectively-acting part of the symmetry is $\Z_2$, differentiated by its $\Z_2$-valued anomaly.\\

With this setup, let us perform some manipulations of the 2d partial traces we would obtain when gauging the $\Z_4$.  Concretely, we select the condensable algebras $\mathcal{B}_{\text{sym.}}=1\oplus e\oplus e^2\oplus e^3$ and $\mathcal{B}_{\text{phys.}}=1\oplus e^2m^2$ to define boundary conditions for our SymTFT.  We follow the same process as \cite[Section 3.2]{Robbins:2022wlr}, but this time we regard the TPOs living on the $m$ lines as connected to bulk anyons.  There are two TPOs, one of which is the identity operator and the other which fuses as $\Z_2$.  Since the full theory does not exhibit ground state degeneracy, the non-trivial TPO must stem from a non-trivial anyon which is condensed on the physical bondary, which can only be $e^2m^2$.

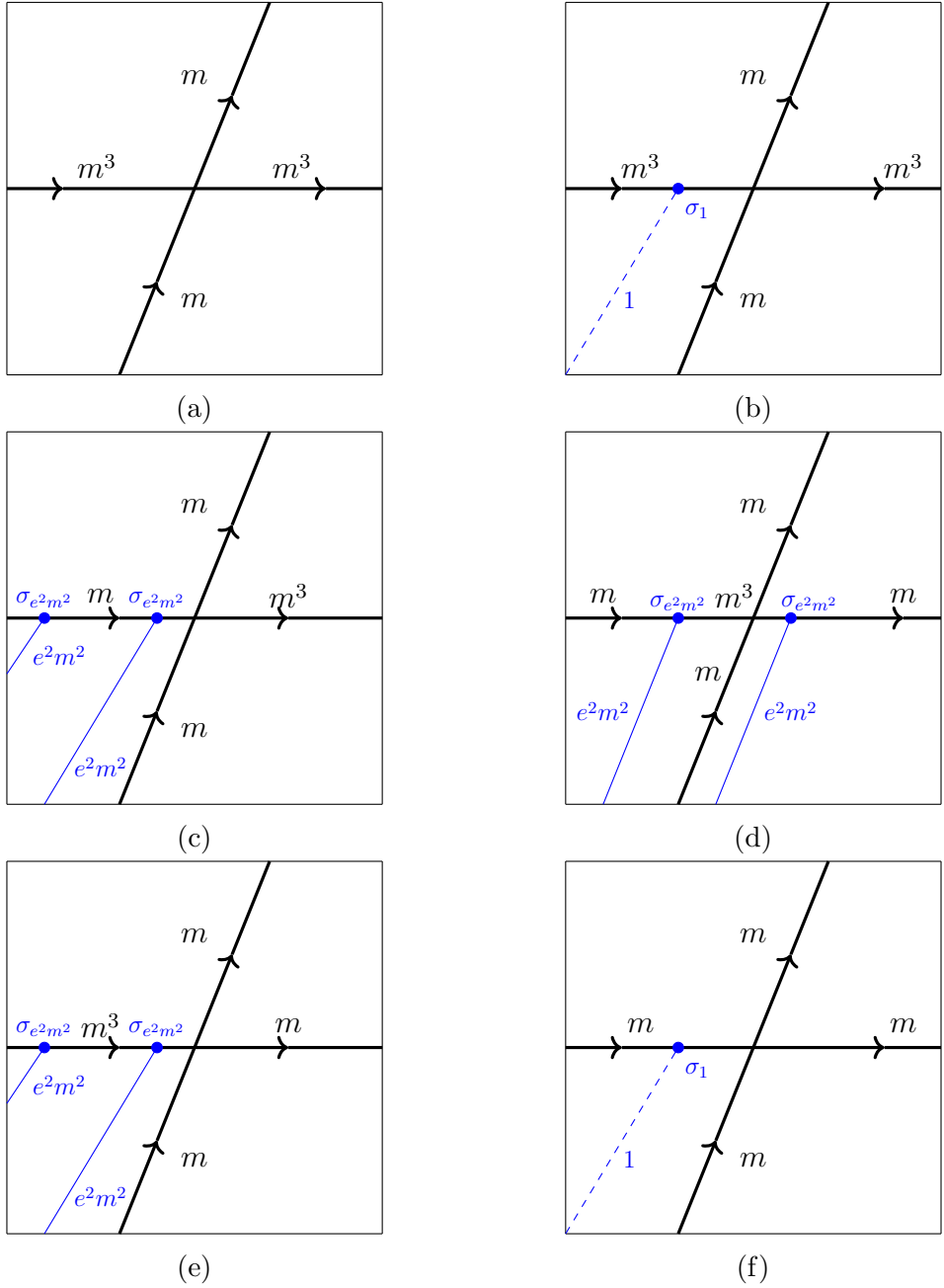
\begin{figure}[p]
	\begin{subfigure}{0.5\textwidth}
		\centering
		\begin{tikzpicture}
		\draw[thin] (0,0)--(5,0);
		\draw[thin] (5,0)--(5,5);
		\draw[thin] (0,0)--(0,5);
		\draw[thin] (0,5)--(5,5);
		\draw[very thick,->] (1.5,0)--(2,1.25);
		\draw[very thick] (2,1.25)--(2.5,2.5);
		\draw[very thick,->] (2.5,2.5)--(3,3.75);
		\draw[very thick] (3,3.75)--(3.5,5);
		\draw[very thick] (4.25,2.5)--(5,2.5);
		\draw[very thick,->] (2.5,2.5)--(4.25,2.5);
		\draw[very thick] (2.5,2.5)--(1.5,2.5);
		\draw[very thick] (0.75,2.5)--(1.5,2.5);
		\draw[very thick,->] (0,2.5)--(0.75,2.5);
		\node at (2.5,4) {$m$};
		\node at (2.5,1) {$m$};
		\node at (1.2,2.8) {$m^3$};
		\node at (3.8,2.8) {$m^3$};
		\end{tikzpicture}
		\caption{}
		\label{fig:pta}
	\end{subfigure}
	\begin{subfigure}{0.5\textwidth}
		\centering
		\begin{tikzpicture}
		\draw[thin] (0,0)--(5,0);
		\draw[thin] (5,0)--(5,5);
		\draw[thin] (0,0)--(0,5);
		\draw[thin] (0,5)--(5,5);
		\draw[very thick,->] (1.5,0)--(2,1.25);
		\draw[very thick] (2,1.25)--(2.5,2.5);
		\draw[very thick,->] (2.5,2.5)--(3,3.75);
		\draw[very thick] (3,3.75)--(3.5,5);
		\draw[very thick] (4.25,2.5)--(5,2.5);
		\draw[very thick,->] (2.5,2.5)--(4.25,2.5);
		\draw[very thick] (2.5,2.5)--(1.5,2.5);
		\draw[very thick] (0.75,2.5)--(1.5,2.5);
		\draw[very thick,->] (0,2.5)--(0.75,2.5);
		\node at (2.5,4) {$m$};
		\node at (2.5,1) {$m$};
		\node at (1,2.8) {$m^3$};
		\node at (4.5,2.8) {$m^3$};
		\filldraw[blue] (1.5,2.5) circle (2pt);
		\node[blue,scale=0.8] at (1.75,2.2) {$\sigma_1$};
        \draw[dashed, blue] (0,0)--(1.5,2.5);
        \node[blue,scale=0.8] at (0.85,1) {$1$};
		\end{tikzpicture}
		\caption{}
		\label{fig:ptb}
	\end{subfigure}
	\begin{subfigure}{0.5\textwidth}
		\centering
		\begin{tikzpicture}
		\draw[thin] (0,0)--(5,0);
		\draw[thin] (5,0)--(5,5);
		\draw[thin] (0,0)--(0,5);
		\draw[thin] (0,5)--(5,5);
		\draw[very thick,->] (1.5,0)--(2,1.25);
		\draw[very thick] (2,1.25)--(2.5,2.5);
		\draw[very thick,->] (2.5,2.5)--(3,3.75);
		\draw[very thick] (3,3.75)--(3.5,5);
		\draw[very thick] (3.75,2.5)--(5,2.5);
		\draw[very thick,->] (2.5,2.5)--(3.75,2.5);
		\draw[very thick] (2.5,2.5)--(1.5,2.5);
		\draw[very thick,->] (0,2.5)--(1.5,2.5);
		\node at (2.5,4) {$m$};
		\node at (2.5,1) {$m$};
		\node at (1.25,2.8) {$m$};
		\node at (3.75,2.8) {$m^3$};
		\filldraw[blue] (2,2.5) circle (2pt);
		\node[blue,scale=0.8] at (2,2.75) {$\sigma_{e^2m^2}$};
        \draw[blue] (0.5,0)--(2,2.5);
        \node[blue,scale=0.8] at (0.70,2) {$e^2m^2$};
		\filldraw[blue] (0.5,2.5) circle (2pt);
		\node[blue,scale=0.8] at (0.5,2.75) {$\sigma_{e^2m^2}$};
        \draw[blue] (0,1.75)--(0.5,2.5);
        \node[blue,scale=0.8] at (1.25,0.5) {$e^2m^2$};
		\end{tikzpicture}
		\caption{}
		\label{fig:ptc}
	\end{subfigure}
	\begin{subfigure}{0.5\textwidth}
		\centering
		\begin{tikzpicture}
		\draw[thin] (0,0)--(5,0);
		\draw[thin] (5,0)--(5,5);
		\draw[thin] (0,0)--(0,5);
		\draw[thin] (0,5)--(5,5);
		\draw[very thick,->] (1.5,0)--(2,1.25);
		\draw[very thick] (2,1.25)--(2.5,2.5);
		\draw[very thick,->] (2.5,2.5)--(3,3.75);
		\draw[very thick] (3,3.75)--(3.5,5);
		\draw[very thick] (4.5,2.5)--(5,2.5);
		\draw[very thick,->] (2.5,2.5)--(4.5,2.5);
		\draw[very thick] (2.5,2.5)--(1.5,2.5);
		\draw[very thick] (0.75,2.5)--(1.5,2.5);
		\draw[very thick,->] (0,2.5)--(0.75,2.5);
		\node at (2.5,4) {$m$};
		\node at (1.9,1.75) {$m$};
		\node[scale=1] at (2.25,2.8) {$m^3$};
		\node at (0.5,2.8) {$m$};
		\node at (4.5,2.8) {$m$};
		\filldraw[blue] (1.5,2.5) circle (2pt);
		\node[blue,scale=0.8] at (1.5,2.7) {$\sigma_{e^2m^2}$};
        \draw[blue] (0.5,0)--(1.5,2.5);
        \node[blue,scale=0.8] at (0.5,1.25) {$e^2m^2$};
		\filldraw[blue] (3.0,2.5) circle (2pt);
		\node[blue,scale=0.8] at (3.25,2.7) {$\sigma_{e^2m^2}$};
        \draw[blue] (2.0,0)--(3.0,2.5);
        \node[blue,scale=0.8] at (3,1.25) {$e^2m^2$};
		\end{tikzpicture}
		\caption{}
		\label{fig:ptd}
	\end{subfigure}
	\begin{subfigure}{0.5\textwidth}
		\centering
		\begin{tikzpicture}
		\draw[thin] (0,0)--(5,0);
		\draw[thin] (5,0)--(5,5);
		\draw[thin] (0,0)--(0,5);
		\draw[thin] (0,5)--(5,5);
		\draw[very thick,->] (1.5,0)--(2,1.25);
		\draw[very thick] (2,1.25)--(2.5,2.5);
		\draw[very thick,->] (2.5,2.5)--(3,3.75);
		\draw[very thick] (3,3.75)--(3.5,5);
		\draw[very thick] (3.75,2.5)--(5,2.5);
		\draw[very thick,->] (2.5,2.5)--(3.75,2.5);
		\draw[very thick] (2.5,2.5)--(1.5,2.5);
		\draw[very thick,->] (0,2.5)--(1.5,2.5);
		\node at (2.5,4) {$m$};
		\node at (2.5,1) {$m$};
		\node at (1.25,2.8) {$m^3$};
		\node at (3.75,2.8) {$m$};
		\filldraw[blue] (2,2.5) circle (2pt);
		\node[blue,scale=0.8] at (2,2.75) {$\sigma_{e^2m^2}$};
        \draw[blue] (0.5,0)--(2,2.5);
        \node[blue,scale=0.8] at (0.70,2) {$e^2m^2$};
		\filldraw[blue] (0.5,2.5) circle (2pt);
		\node[blue,scale=0.8] at (0.5,2.75) {$\sigma_{e^2m^2}$};
        \draw[blue] (0,1.75)--(0.5,2.5);
        \node[blue,scale=0.8] at (1.25,0.5) {$e^2m^2$};
		\end{tikzpicture}
		\caption{}
		\label{fig:pte}
	\end{subfigure}
	\begin{subfigure}{0.5\textwidth}
		\centering
		\begin{tikzpicture}
		\draw[thin] (0,0)--(5,0);
		\draw[thin] (5,0)--(5,5);
		\draw[thin] (0,0)--(0,5);
		\draw[thin] (0,5)--(5,5);
		\draw[very thick,->] (1.5,0)--(2,1.25);
		\draw[very thick] (2,1.25)--(2.5,2.5);
		\draw[very thick,->] (2.5,2.5)--(3,3.75);
		\draw[very thick] (3,3.75)--(3.5,5);
		\draw[very thick] (4.25,2.5)--(5,2.5);
		\draw[very thick,->] (2.5,2.5)--(4.25,2.5);
		\draw[very thick] (2.5,2.5)--(1.5,2.5);
		\draw[very thick] (0.75,2.5)--(1.5,2.5);
		\draw[very thick,->] (0,2.5)--(0.75,2.5);
		\node at (2.5,4) {$m$};
		\node at (2.5,1) {$m$};
		\node at (1,2.8) {$m$};
		\node at (4.5,2.8) {$m$};
		\filldraw[blue] (1.5,2.5) circle (2pt);
		\node[blue,scale=0.8] at (1.75,2.2) {$\sigma_1$};
        \draw[dashed, blue] (0,0)--(1.5,2.5);
        \node[blue,scale=0.8] at (0.85,1) {$1$};
		\end{tikzpicture}
		\caption{}
		\label{fig:ptf}
	\end{subfigure}
	\caption{We map the $Z_{m,m^3}$ partial trace to $Z_{m,m}$ by regarding it as living on the boundary of a SymTFT, shown above.  Black lines lie on the boundary.
    Blue lines extend into the bulk.  The dashed line is the identity operator.  Note in going from diagram (d) to (e), one gets a sign.}
	\label{fig:pt}
\end{figure}

We illustrate in Figure~\ref{fig:pt} how the partial trace $Z_{m,m^3}$ is mapped to $Z_{m,m}$.  Beginning in Figure~\ref{fig:pta}, we illustrate the topological lines of $Z_{m,m^3}$, which in the SymTFT should be thought of as living on the symmetry boundary.  In Figure~\ref{fig:ptb} we add the local identity operator on the $m^3$ line, connected to the identity anyon.\footnote{Bulk lines and the boundary operators to which they connect are drawn in blue.}  Moving to Figure~\ref{fig:ptc} we split this line/operator into two pieces: $e^2m^2$ lines ending on non-trivial TPOs labeled $\sigma_{e^2m^2}$.  This splitting is simply the inverse of their $\Z_2$ fusion.  Since this space is meant to be a torus (the top/bottom and left/right boundaries are identified), we get to Figure~\ref{fig:ptd} simply by moving the previous figure's leftmost TPO farther left, wrapping it around the cycle.  The transition from Figure~\ref{fig:ptd} to Figure~\ref{fig:pte} is the most important, as it involves pulling the $\sigma_{e^2m^2}$ operator across the vertical $m$ line.  Alternatively, we could get the same result by pulling the vertical $m$ line slightly off the boundary, moving the $e^2m^2$ line past it, and then putting it back.  That makes it clear that whatever phase is generated here comes from the braiding of the anyons labeled by $m$ and $e^2m^2$.  For anyons in the Drinfeld center of an abelian group, braiding is given simply by characters, so we can immediately get the phase incurred by this swap as $\chi_{e^2}(m)=-1$ (regarding $m$ as generating $\Z_4$ and $e$ as generating $\hat{\Z}_4$).  Finally, Figure~\ref{fig:ptf} follows simply by fusing the lines once again.  The remaining identity operators can be erased and we are left with the partial trace $Z_{m,m}$.  Taking into account the braiding phase that we accrued along the way, we conclude that $Z_{m,m^3}$ is equal to $-Z_{m,m}$, exactly as we claimed based on pure boundary calculations earlier.  Note as well that the other possible boundary condition which would have given the $\Z_4$ a trivially-acting $\Z_2$ center, $\mathcal{B}_{\text{phys.}}=1\oplus m^2$, would have led to the non-trivial $\Z_2$ TPO on the symmetry boundary being connected to the $m^2$ anyon.  Since all of the $m$ lines braid trivially with each other, we would not have found any additional phases entering the partial traces here, consistent with the fact that the effectively-acting $\Z_2$ in this case is not anomalous.

\subsection{Incorporating trivial symmetries directly in SymTFT}\label{sec:trivial_sym}

In the previous subsection we presented a realization of anomaly resolution in SymTFT, in which two bulk phases, for the anomalous effective symmetry and for the non-anomalous extended symmetry, are separated by a topological interface.  The extended symmetry lines which end on the interface, i.e.~which don't correspond to anything in the effective symmetry phase, represent trivially acting symmetries.  Nothing that is attached to the physical boundary will be charged under them.

There is an alternative approach one can take~\cite{Lin:2025oml}, which is a bit of a departure from the usual SymTFT, in which we have just a single bulk phase incorporating the extended symmetry, and the trivially acting symmetries are implemented by ensuring that those lines can end on topological point operators.  Such lines can always be opened up to separate from any linking with other lines, which means that nothing can be charged under them; they are trivially acting.

Note that in this subsection we use a description in terms of continuous differential forms rather than the discrete forms which appear in the last subsection.

Recall that the 3d SymTFT associated with a 2d QFT $\mathcal{T}$ possessing a $\Z_{2_e}$ symmetry is described by the action: 
\begin{equation}\label{eqn:Z2EffSym}
    S_{bulk} = \frac{i}{\pi} \int a_1 d b_1 .
\end{equation}
The topological operators in the bulk are given by 
\begin{equation}\label{eqn:Z2_Op}
    Y[\gamma_1] = e^{i \int_{\gamma_1} a_1}, \qquad X[\gamma'_1] = e^{i \int_{\gamma'_1} b_1},
\end{equation}
where $\gamma_1$ and $\gamma'_1$ are 1-cycles in the bulk.

In addition to the effective $\Z_{2 e}$ symmetry, we would like to introduce a trivially acting $\Z_{2 t}$ symmetry to this theory. This new symmetry can either form a direct product with the existing $\Z_{2 e}$ symmetry resulting in a ${\Z_2}_t \times {\Z_2}_e$ symmetry, or it can mix non-trivially with $\Z_{2 e}$ and form a extended $\Z_4$ symmetry group acting on $\mathcal{T}$. The resulting symmetry group depends on the choice of extension class.

\subsubsection*{Case I : Trivial extension}

We begin by considering the trivial extension. 
\begin{equation}
    0 \longrightarrow {\Z_2}_t \longrightarrow {\Z_2}_t \times {\Z_2}_e \longrightarrow {\Z_2}_e \longrightarrow 0 .
\end{equation}
We have labeled the $\Z_2$ factors with subscripts to distinguish them. Starting from a theory with a $\Z_{2 e}$ symmetry, we can promote it to a theory with ${\Z_2}_t \times {\Z_2}_e$ symmetry by introducing additional couplings to the SymTFT action in equation \eqref{eqn:Z2EffSym}. The modified SymTFT action, which captures this ${\Z_2}_t \times {\Z_2}_e$ symmetry, is 
\begin{equation}
     S_{bulk} = \frac{i}{\pi} \int \Bigl( a_1d b_1 + \underbrace{\tilde{a}_1d\tilde{b}_1 + c_2db_0 + c_2\tilde{b}_1}_{\text{trivial} ~ \Z_2} \Bigr),
\end{equation}
where the last three couplings take care of the $\Z_{2 t}$. This action is gauge invariant under the following gauge transformations:
\begin{equation}\label{eqn:gauge_tran}
    \begin{aligned}
        a_1 &\longrightarrow a_1 + d\lambda_0, \\
        b_1 &\longrightarrow b_1 + d\mu_0, \\
        \tilde{a}_1 &\longrightarrow \tilde{a}_1 + d\tilde{\lambda}_0 - \mu_1, \\
        \tilde{b}_1 &\longrightarrow \tilde{b}_1 + d\tilde{\mu}_0, \\
        b_0 &\longrightarrow b_0 -\tilde{\mu}_0, \\
        c_2 &\longrightarrow c_2 +d\mu_1.
    \end{aligned}
\end{equation}
The corresponding equations of motion are
\begin{equation}
        \begin{aligned}
        b_1 &: da_1 = 0, \qquad a_1 : db_1= 0 \\
        \tilde{a}_1 &: d\tilde{b}_1 = 0, \qquad b_0 : dc_2 = 0, \\
        \tilde{b}_1 &: d\tilde{a}_1 + c_2 = 0, \qquad
        c_2: db_0 + \tilde{b}_1 = 0  .  
        \end{aligned}
\end{equation}
In addition to the line operators in \eqref{eqn:Z2_Op}, the additional couplings give rise to new topological operators in the bulk, given by  
\begin{equation}
\begin{aligned}
    \tilde{X}[\partial \delta_1 = y, \delta_1] &= \text{exp}\left(i\int_{\delta_1} \tilde{b}_1 + i b_0(y) \right), \\ 
    V[\gamma_1 = \partial \Sigma_2, \Sigma_2] &=\exp\left(i\int_{\Sigma_2}c_2 + i\oint_{\gamma_1}\tilde{a}_1\right).
 \end{aligned}
\end{equation}
We observe that the SymTFT is engineered in a way, there are no free topological point operators in the bulk. We now turn to a discussion of various boundary conditions (summarized in Table~\eqref{table:trivial_ext}) of this SymTFT. 
\begin{itemize}
    \item We realize $\Z_{2 t} \times \Z_{2 e}$ symmetry by imposing the Dirichlet boundary condition on $a_1, \tilde{a}_1, c_2$ and Neumann boundary condition on $b_1, \tilde{b}_1, b_0$. Under these conditions, the line operator $\tilde{X}[\partial \delta_1 = y, \delta_1]$ remains parallel to the symmetry boundary and  generates the trivially acting $\Z_{2 t}$ symmetry. Since they end on topological point operators, they do not link with any other operators. In contrast, the $X[\gamma'_1]$ line operators generate effective $\Z_{2 e}$ symmetry. 
    \item We take Dirichlet b.c. for $b_1, \tilde{b}_1,c_2$ and Neumann b.c for $a_1, \tilde{a}_1,b_0$, this corresponds to gauging $\Z_{2 t} \times \Z_{2 e}$. This gauging results in free topological point operators on the symmetry boundary, leading to a decomposing theory. The surface operator $V[\gamma_1 = \partial \Sigma_2, \Sigma_2]$ terminates on line operators, which remain topological on the symmetry boundary. These line operators act as domain walls, separating distinct universes.
    
    As a special case, one may instead gauge a subgroup of $\Z_{2 t} \times \Z_{2 e}$. Gauging the trivially acting subgroup still leads to decomposition. In contrast, gauging the effective $\Z_{2 e}$ does not result in a decomposing theory. 
    \item Finally, we can undo the decomposition by gauging the 1-form symmetry. In order to that we need to impose Dirichlet boundary condition on $b_1, \tilde{b}_1,b_0$ and Neumann on $a_1, \tilde{a}_1,c_2$. The surface operator $V[\gamma_1 = \partial \Sigma_2, \Sigma_2]$ will now generate a $(-1)$-form symmetry.   
\end{itemize}
\begin{table}[H]
\centering
\begin{tabular}{|c|c|c|}
    \hline
    \textbf{Cases} & \textbf{Dirichlet} & \textbf{Neumann} \\
    \hline
    ${\Z_2}_t \times {\Z_2}_e$ & $a_1, \tilde{a}_1,c_2$  & $b_1, \tilde{b}_1,b_0$ \\
    \hline
    Gauging ${\Z_2}_t \times {\Z_2}_e$ and Decomposition & $b_1, \tilde{b}_1,c_2$ & $a_1, \tilde{a}_1,b_0$  \\
    \hline
    Gauging ${\Z_2}_t \subset {\Z_2}_t \times {\Z_2}_e$ and Decomposition & $a_1, \tilde{b}_1, c_2$ & $\tilde{a}_1, b_0,b_1$ \\
    \hline
    Gauging ${\Z_2}_e \subset {\Z_2}_t \times {\Z_2}_e$ & $b_1, \tilde{a}_1, c_2$ & $a_1 ,\tilde{b}_1, b_0$ \\
    \hline 
    Undoing Decomposition & $b_1, \tilde{b}_1,b_0$ & $a_1, \tilde{a}_1,c_2$ \\
    \hline
\end{tabular}
\caption{\footnotesize{Trivial extension}}\label{table:trivial_ext}
\end{table}

\subsubsection*{Case II : Non-trivial extension}

We move on to discuss the case where the two $\Z_2$'s mix non-trivially to form a $\Z_4$: 
\begin{equation}
    0 \longrightarrow {\Z_2}_t \longrightarrow \Z_4 \longrightarrow {\Z_2}_e \longrightarrow 0.
\end{equation}
In order to get a $\Z_4$, we need an additional coupling in the SymTFT action, 
\begin{equation}\label{eqn:S_bulk_w/o_anom}
     S_{bulk} = \frac{i}{\pi} \int  \left( a_1 d b_1 + \tilde{a}_1d\tilde{b}_1 + c_2db_0 + c_2\tilde{b}_1 - \frac{1}{2} a_1 d\tilde{b}_1 \right).
\end{equation}
The action is gauge invariant under \eqref{eqn:gauge_tran} and the local equations of motion are, 
\begin{equation}
    \begin{aligned}
        \tilde{a}_1&: ~ d\tilde{b}_1  =0,  \\
        b_1&: ~ da_1 = 0, \\
        b_0&: ~ dc_2 = 0, \\
        c_2&: ~ db_0 + \tilde{b}_1 =0, \\
        a_1&: ~ db_1 - \frac{1}{2} d\tilde{b}_1 = 0 ,\\
        \tilde{b}_1&: ~ d\tilde{a}_1 - \frac{1}{2} da_1 + c_2 = 0.
    \end{aligned}
\end{equation}
We acquire a QFT with $\Z_4$ symmetry (with a trivially acting subgroup) by imposing the Dirichlet boundary condition on $a_1, \tilde{a}_1, c_2$ and the Neumann boundary condition on $b_1, \tilde{b}_1, b_0$. Like before, we have a topological line operator that ends on a point operator in the bulk, that generates the trivially acting subgroup $\Z_{2 t} \subset \Z_4$.

To see that it is in fact $\Z_4$, note that summing over the $a_1$ gauge bundle degrees of freedom implies that $2\oint b_1-\oint\tilde{b}_1\in 2\pi\Z$, and this implies that a $\tilde{X}$ operator defined on a loop, $\tilde{X}[\phi,\gamma_1]=\exp(i\oint_{\gamma_1}\tilde{b}_1)$, will satisfy $\tilde{X}^2=X$.  Since $X^2=1$, this $\tilde{X}$ operator is a generator of a $\Z_4$ symmetry.

We can gauge the $\Z_4$ symmetry by changing the boundary condition on the symmetry boundary, produces genuine topological point operators in the bulk. The fact that $\Z_2$'s mix non-trivially is corroborated from the fact that we can no longer gauge $\Z_{2 e}$ like in case I. We can try to impose Dirichlet b.c. on $b_1, \tilde{a}_1, c_2$ and Neumann b.c. on $a_1, \tilde{b}_1, b_0$, which corresponds to gauging $\Z_{2 e}$ but this particular choice of boundary conditions is no longer allowed. In order to see that we vary the action, 
\begin{equation}
    \begin{aligned}
        \delta S_{bulk} = \int_{M_3}\Bigl(  \delta a_1db_1 + \textcolor{blue}{a_1d\delta b_1} + \delta \tilde{a}_1d\tilde{b}_1 + \textcolor{blue}{\tilde{a}_1d\delta\tilde{b}_1} \\ + \delta c_2db_0 + \textcolor{blue}{c_2 d\delta b_0} + \delta c_2 \tilde{b}_1 + c_2 \delta \tilde{b}_1 \\
        - ~ \frac{1}{2} \delta a_1 d\tilde{b}_1 - \textcolor{blue}{\frac{1}{2} a_1 d\delta\tilde{b}_1} \Bigr).
    \end{aligned}
\end{equation}
Doing an integration by parts on the blue colored terms, 
\begin{equation}
    \begin{aligned}
        \delta S_{bulk} = \int_{M_3} \Bigl( \delta a_1db_1 + \delta \tilde{a}_1d\tilde{b}_1 + \delta c_2db_0 + \delta c_2 \tilde{b}_1 + c_2 \delta \tilde{b}_1 - \\ \frac{1}{2} \delta a_1 d\tilde{b}_1 - da_1\delta b_1 - d\tilde{a}_1\delta \tilde{b}_1 - dc_2 \delta b_0 + \frac{1}{2} da_1\delta \tilde{b}_1 \Bigr) \\
        + ~ \textcolor{gray}{ \int_{\partial M_3} \Bigl( a_1\delta b_1 + \tilde{a}_1 \delta \tilde{b}_1 + c_2\delta b_0 - \frac{1}{2} a_1 \delta \tilde{b}_1 \Bigr) }.
    \end{aligned}
\end{equation}
The gray-colored terms must vanish to have a well-posed variational problem.  We can clearly see that we cannot impose Neumann boundary condition on $a_1, \tilde{b}_1,b_0$ as the last term would not vanish, also we cannot remove that by adding any boundary term, confirming the fact that $\Z_{2_e}$ is not a subgroup of $\Z_4$. We tabulate the list of relevant boundary conditions,
\begin{table}[H]
\centering
\begin{tabular}{|c|c|c|}
    \hline
    Allowed B.C & Dirichlet & Neumann \\
    \hline
    $\Z_4$ with Trivial $\Z_2$ & $a_1, \tilde{a}_1,c_2$  & $b_1, \tilde{b}_1,b_0$ \\
    \hline
    Gauging $\Z_4$ and Decomposition & $b_1, \tilde{b}_1,c_2$ & $a_1, \tilde{a}_1,b_0$  \\
    \hline
    Gauging ${\Z_2}_t \subset \Z_4$ & $a_1, \tilde{b}_1, c_2$ & $b_1, \tilde{a}_1, b_0$ \\
    \hline 
    Undoing Decomposition & $b_1, \tilde{b}_1,b_0$ & $a_1, \tilde{a}_1,c_2$ \\
    \hline
\end{tabular}
\caption{Non-trivial extension}
\end{table}

\subsubsection*{Case III : Anomaly resolution}

Finally we discuss the case where we resolve the anomaly by introducing additional symmetry. We start with an anomalous $\Z_2$ QFT, the SymTFT for such a theory is given by the Dijkgraaf--Witten theory with a topological twist. 
\begin{equation}
    \begin{aligned}
        S_{bulk} = \frac{i}{\pi} \int  \left( a_1db_1 + \frac{1}{2} a_1da_1 \right).
    \end{aligned}
\end{equation}
The presence of anomalous coupling rules out the Neumann boundary condition for $a_1$. 
If we vary the action, 
\begin{equation*}
    \begin{aligned}
        \delta S = \int_{M_3} \left( \delta a_1 db_1 - da_1 \delta b_1 + \frac{1}{2}\delta a_1 da_1 - \frac{1}{2}da_1 \delta a_1 \right) +\int_{\partial M_3} \left (\frac{1}{2} a_1 \delta a_1 + a_1 \delta b_1 \right).
    \end{aligned}
\end{equation*}
To have a well-defined variational problem, we need the boundary terms to vanish. We can clearly see that, if $a_1$ fluctuates on the boundary, we cannot cancel the boundary term $a_1\delta a_1$.  

We intend to resolve the anomaly by adding trivial symmetries. We expect the additional symmetries to mix non-trivially with the existing anomalous symmetry. The situation can be summarized by a short exact sequence as follows, 
\begin{align}
    0 \longrightarrow \Z_{2 t} \longrightarrow \Z_4 \longrightarrow \Z_2^\omega \longrightarrow 0.
\end{align}
As a warm up, we would like to resolve the anomaly by introducing an effective $\Z_2$, then we will introduce the couplings necessary to make this $\Z_2$ trivially acting. So, we begin with, 
\begin{equation}\label{eqn:S_bulk_anom}
    S_{bulk} = \frac{i}{\pi} \int \left( a_1db_1 + \frac{1}{2} a_1da_1 + \tilde{a}_1d\tilde{b}_1 - \frac{1}{2}a_1d\tilde{b}_1 \right).
\end{equation}
The last term in the SymTFT action just couples the $\Z_2$'s. In order to see that the anomaly is truly resolved, we vary the action: 
\begin{equation}
    \begin{aligned}
        \delta S_{bulk} = \int_{M_3} \Bigl( \delta a_1db_1 + \textcolor{blue}{a_1d\delta b_1} + \frac{1}{2} \delta a_1da_1 + \textcolor{blue}{\frac{1}{2} a_1d\delta a_1} + \delta \tilde{a}_1d\tilde{b}_1 \\
        + ~ \textcolor{blue}{\tilde{a}_1d\delta\tilde{b}_1}
        - \frac{1}{2} \delta a_1 d\tilde{b}_1 - \textcolor{blue}{\frac{1}{2} a_1 d\delta\tilde{b}_1} \Bigr).
    \end{aligned}
\end{equation}
Doing an integration by parts on the blue colored terms, 
\begin{equation}
    \begin{aligned}
        \delta S_{bulk} = \int_{M_3} \Bigl( \delta a_1db_1 + \frac{1}{2} \delta a_1da_1 + \delta \tilde{a}_1d\tilde{b}_1 - \frac{1}{2} \delta a_1 d\tilde{b}_1 \\ - ~da_1\delta b_1 -\frac{1}{2} da_1 \delta a_1 - d\tilde{a}_1\delta \tilde{b}_1 + \frac{1}{2} da_1\delta \tilde{b}_1 \Bigr) \\
        + ~  \textcolor{gray}{ \int_{\partial M_3} \Bigl( a_1\delta b_1 + \frac{1}{2}a_1 \delta a_1 + \tilde{a}_1 \delta \tilde{b}_1 - \frac{1}{2} a_1 \delta \tilde{b}_1 \Bigr)} .
    \end{aligned}
\end{equation}
The boundary terms (in gray) must vanish. To achieve that, we add a boundary term $a_1\tilde{a}_1$, which modifies the boundary variation, 
\begin{align*}
    \delta S= \text{bulk terms} ~ + \int_{\partial M_3} \left( a_1\delta b_1 + \frac{1}{2}a_1 \delta a_1 + \tilde{a}_1 \delta \tilde{b}_1 - \frac{1}{2} a_1 \delta \tilde{b}_1 + a_1\delta\tilde{a}_1 - \tilde{a}_1\delta a_1 \right).
\end{align*}
We impose, 
\begin{align}
\label{eq:NewBCs}
    \tilde{b}_1 = a_1| ~ &\leftrightarrow ~\delta\tilde{b}_1 = \delta a_1, \\
    b_1 = -\tilde{a}_1| ~ &\leftrightarrow ~ \delta b_1 = - \delta \tilde{a}_1.
\end{align}
So, the boundary variation vanishes, even when $a_1, \tilde{a}_1$ is fluctuating on the boundary. Hence, we can claim that the anomaly has been resolved. 

Another perspective is that these boundary conditions amount to imposing Dirichlet boundary conditions on redefined fields $b_1'=b_1+\tilde{a}_1$ and $\tilde{b}_1'=\tilde{b}_1-a_1$.  Written in terms of the fields $a_1$, $\tilde{a}_1$, $b_1'$ and $\tilde{b}_1'$, the action \eqref{eqn:S_bulk_anom} simply becomes the action for a non-anomalous effective $\Z_4$ symmetry.

Now we can add the $c_2db_0$ coupling to \eqref{eqn:S_bulk_anom} to make one of the $\Z_2$ act trivially. We end up with, 
\begin{equation}
    S_{bulk} = \frac{i}{\pi} \int \left(  a_1db_1 + \frac{1}{2} a_1da_1 + \tilde{a}_1d\tilde{b}_1 + c_2db_0 + c_2\tilde{b}_1 - \frac{1}{2}a_1d\tilde{b}_1 \right).
\end{equation}
This action is gauge invariant under the following gauge transformations, 
\begin{equation}
    \begin{aligned}
        a_1 &\longrightarrow a_1 + d\lambda_0, \\
        b_1 &\longrightarrow b_1 + d\mu_0, \\
        \tilde{a}_1 &\longrightarrow \tilde{a}_1 + d\tilde{\lambda}_0 - \mu_1, \\
        \tilde{b}_1 &\longrightarrow \tilde{b}_1 + d\tilde{\mu}_0, \\
        b_0 &\longrightarrow b_0 -\tilde{\mu}_0, \\
        c_2 &\longrightarrow c_2 +d\mu_1 .
    \end{aligned}
\end{equation}
The local equation of motions are, 
\begin{equation}
    \begin{aligned}
        b_1: da_1 =0, \quad \tilde{a}_1: d\tilde{b}_1= 0,  \quad b_0: dc_2 =0 \quad c_2: db_0 + \tilde{b}_1 &=0 , \\
        \tilde{b}_1: d\tilde{a}_1 - \frac{1}{2} da_1 + c_2 =0, \quad a_1: db_1 + da_1 - \frac{1}{2} d\tilde{b}_1 &=0 . \\
    \end{aligned}
\end{equation}
Once again, we realize the line operators that end on point operators in the bulk. 

If we now want to gauge the full $\Z_4$, we can do it by imposing boundary conditions \eqref{eq:NewBCs}, along with Dirichlet conditions for $c_2$ and Neumann conditions for $b_0$.  However, unlike the non-anomalous case, we do not get decomposition in this situation.  The reason is that the $b_0$ point operators sit at the ends of $\tilde{X}$ lines which no longer vanish at the boundary.  Instead, $\tilde{X}$ is essentially $\tilde{X}'Y$.  At the boundary $\tilde{X}'$ vanishes, but $Y$ does not, so the $b_0$ point operators are not genuine.

\begin{table}[H]
\centering
\begin{tabular}{|c|c|c|}
    \hline
    Allowed B.C & Dirichlet & Neumann \\
    \hline
    $\Z_4$ with Trivial ${\Z_2}_t$ & $a_1, \tilde{a}_1,c_2$  & $b_1, \tilde{b}_1,b_0$ \\
    \hline
    Gauging $\Z_4$ & $b_1, \tilde{b}_1,c_2$  & $a_1, \tilde{a}_1,b_0$  \\
    \hline 
    Gauging ${\Z_2}_t \subset \Z_4$ & $a_1, \tilde{b}_1, c_2$ & $b_1, \tilde{a}_1,b_0$  \\
    \hline
\end{tabular}
\caption{Anomaly resolution}
\end{table}

\subsection{General picture in quasi-Hopf algebras}
\label{sect:z2:z4:mixedanoom}

Next, we turn to the algebraic approach.  We describe the original global symmetry $G$ with 't Hooft anomaly $\alpha \in H^3(G,U(1))$ in terms of a quasi-Hopf algebra $H$ such that
\begin{equation}
    {\rm Rep}(H) \: \cong \: {\rm Vec}(G,\alpha).
\end{equation}
In this fashion, the quasi-Hopf algebra implicitly encodes the anomaly.  Then, the idea of the resolution is to find fusion categories ${\cal C}$, ${\cal K}$ for which there is an exact sequence
\begin{equation}
    {\cal K} \: \longrightarrow \: {\cal C} \: \longrightarrow \: {\rm Vec}(G,\alpha)
\end{equation}
and for which ${\cal C}$ admits a fiber functor.  Then, ${\cal C}$ acts as a resolution of
Vec$(G,\alpha)$.  In principle, this will coincide with the group-like case if ${\cal C} = {\rm Vec}(\Gamma)$ and ${\cal K} = {\rm Vec}(K)$.  

Next, we turn to the details of differentials and so forth.

Suppose we have a situation as described in Section~\ref{subsec:GroupGeneralPicture}, in which a group $G$ with an anomaly encoded by a cocycle $\alpha\in H^3(G,U(1))$ is extended by trivially acting symmetries $K$ to a group $\Gamma$,
\begin{equation}
    1\longrightarrow K\longrightarrow\Gamma\stackrel{\pi}{\longrightarrow}G\longrightarrow 1,
\end{equation}
such that the pullback $[\pi^\ast\alpha]$ of the anomaly
class is trivial in $H^3(\Gamma,U(1))$, i.e.~there exists a 2-cochain $j\in C^2(\Gamma,U(1))$ such that $dj=\pi^\ast\alpha$.  We would now like to describe how this looks from the categorical point of view, where we have a strong monoidal functor $(F,J)$ from ${\rm Vec}(\Gamma)$ to ${\rm Vec}(G,\alpha)$.

First we recall that the simple objects in ${\rm Vec}(G,\alpha)$ are labeled by elements $g\in G$ and their fusion is given by the group multiplication, $g_1\otimes g_2=(g_1g_2)$.  The Hom spaces between simple objects ${\rm Hom}(g,h)$ are either zero\footnote{In fact, when we have a normal subgroup $K\subset\Gamma$ that acts trivially, then this is no longer true; instead the space ${\rm Hom}_\Gamma(\gamma_1,\gamma_2)$ is isomorphic to $\C$ whenever $\gamma_1\gamma_2^{-1}\in K$.  This fact is important in explaining decomposition in such cases, since it is point operators at the end of $K$ lines (i.e.~living in ${\rm Hom}(k,1)$) that become the topological point operators heralding decomposition once we gauge $K$~\cite{Robbins:2022wlr}.  However, we don't need to worry about such extra homomorphisms here when we construct the strong monoidal structure.  If $\gamma_1\ne\gamma_2$ then we necessarily have $F(\varphi)=0$ for all $\varphi\in{\rm Hom}(\gamma_1,\gamma_2)$.} if $g\ne h$ or are isomorphic to $\C$ if $g=h$.  In the latter case we have a canonical basis vector $\varphi_g\in{\rm Hom}(g,g)$.  The associators of ${\rm Vec}(G,\alpha)$, a collection of morphisms $a_{g,h,\ell}\in{\rm Hom}((g\otimes h)\otimes\ell,g\otimes(h\otimes\ell))$, are given in terms of the cocycle $\alpha\in Z^3(G,U(1))$ as $a_{g,h,\ell}=\alpha(g,h,\ell)\varphi_{gh\ell}$.

Now to construct a strong monoidal functor from a tensor category $\mathcal{C}$ to another tensor category $\mathcal{D}$, we need to specify a functor $F$ from $\mathcal{C}$ to $\mathcal{D}$, along with a collection of morphisms $J_{X,Y}\in{\rm Hom}_{\mathcal{D}}(F(X)\otimes F(Y),F(X\otimes Y))$ that makes the following diagram commute,
\begin{equation}   \label{eq:relate-assoc}
\begin{tikzcd}
(F(X)\otimes F(Y))\otimes F(Z) \arrow[rrr, "{a_{F(X),F(Y),F(Z)}}"] \arrow[d, "{J_{X,Y}\otimes \text{id}_{F(Z)}}"'] &  &  & F(X)\otimes (F(Y)\otimes F(Z)) \arrow[d, "{\text{id}_{F(X)}\otimes J_{Y,Z}}"] \\
F(X\otimes Y)\otimes F(Z) \arrow[d, "{J_{X\otimes Y,Z}}"']                                                         &  &  & F(X)\otimes F(Y\otimes Z) \arrow[d, "{J_{X,Y\otimes Z}}"]                     \\
F((X\otimes Y)\otimes Z) \arrow[rrr, "{F(a_{X,Y,Z})}"']                                                            &  &  & F(X\otimes (Y\otimes Z))                                                     
\end{tikzcd}
\end{equation}

For the case at hand, we take the functor to act on objects as $F(\gamma)=\pi(\gamma)$, and to act on morphisms as $F(\varphi_\gamma)=\varphi_{\pi(\gamma)}$.  The morphisms $J_{\gamma_1,\gamma_2}$ are constructed in terms of the 2-cochain $j$ as
\begin{equation}
    J_{\gamma_1,\gamma_2}=j(\gamma_1,\gamma_2)^{-1}\varphi_{\pi(\gamma_1\gamma_2)}.
\end{equation}
The commutativity of the diagram (we assume that the associator in ${\rm Vec}(\Gamma)$ is trivial, $a_{\gamma_1,\gamma_2,\gamma_3}=\varphi_{\gamma_1\gamma_2\gamma_3}$) becomes
\begin{equation}
    \alpha(\pi(\gamma_1),\pi(\gamma_2),\pi(\gamma_3))j(\gamma_2,\gamma_3)^{-1}j(\gamma_1,\gamma_2\gamma_3)^{-1}=j(\gamma_1,\gamma_2)^{-1}j(\gamma_1\gamma_2,\gamma_3)^{-1},
\end{equation}
which is simply the group cohomology statement that $dj=\pi^\ast\alpha$.

\subsection{Anomalous $\mathbb{Z}_2$ extended to $\mathbb{Z}_4$ via quasi-Hopf algebra}

\subsubsection{Algebraic approach}\label{sssec:algebraic}

Algebraically, we can understand the anomaly resolution as follows.
The anomalous ${\mathbb Z}_2$ is described by the fusion category
Vec(${\mathbb Z}_2, [\alpha])$, where $[\alpha] \in H^3({\mathbb Z}_2,U(1))$ encodes
the anomaly.  As discussed in section~\ref{sect:genl:quasihopf},
to resolve the anomaly, we pick another fusion category ${\cal C}$, which
admits a fiber functor (hence a special symmetric Frobenius algebra object), and so is gaugeable,
together with a functor
\begin{equation}
    {\cal C} \: \longrightarrow \: {\rm Vec}({\mathbb Z}_2, [\alpha]).
\end{equation}

To describe the anomaly resolution, 
we will describe Vec$({\mathbb Z}_2, [\alpha])$ as the representation category of
a quasi-Hopf algebra
 which we denote as $H(2)$, not gauge equivalent to a Hopf algebra.

The quasi-Hopf algebra $H(2):=\C_{\alpha}^{\Z_2}$ is an example of the algebras $\C^G_{\alpha}$ described in section~\ref{sect:genl:quasihopf}, with $G=\Z_2$ and $\alpha$ a 3-cocycle belonging to the unique nontrivial class in $H^3(\Z_2,U(1))$. The bialgebra structure is that of the group algebra $\C[\Z_2]$, and thus is generated by group-like elements $\{1,g\}$ such that $g^2=1$. We introduce a nontrivial coassociator
\begin{equation}
    \Phi \: = \:  1\otimes 1\otimes 1 - 2p_-\otimes p_-\otimes p_-,
\end{equation}
where $p_-=\tfrac{1}{2}(1-g)$, an antipode $S(g)=g$, and distinguished elements $\alpha=g$, $\beta=1$. Since $H(2)$ is not twist equivalent to a Hopf algebra, the fusion category $\text{Rep}(H(2))$ does not admit a fiber functor. This representation category 
\begin{equation}
    \text{Rep}(H(2)) \: \cong \:  \text{Vec}(\Z_2,[\alpha])
\end{equation}
is the familiar anomalous fusion category $\text{Vec}(\Z_2,[\alpha])$ where $\alpha$ is a 3-cocycle representing the unique nontrivial cohomology class in $H^3(\Z_2,U(1))=\Z_2$.

We now explicitly describe the anomaly resolution
\begin{equation}
   1 \: \longrightarrow \: \Z_2 \: \longrightarrow \: \Z_4 \: \longrightarrow \:  \Z_2 \: \longrightarrow \: 1,
\end{equation}
in this language. As outlined before, the goal is to exhibit a normal inclusion of $H(2)$ into a gauge twist of the Hopf algebra $\C[\Z_4]$.

We denote the generator of the $\Z_4$ by $z$. The relevant gauge twist in this case is
\begin{equation}\label{eq:z4gaugetwist}
    F= \sum_{g,h\in\Z_4}\omega(g,h) \,p_g\otimes p_h = \tfrac{1}{2}(1+z^2)\otimes 1+ \tfrac{1}{2}(1-z^2)\otimes z^2,
\end{equation}
where we use the idempotents
\begin{eqnarray}
    p_1 = \tfrac{1}{4}(1+z+z^2+z^3), & p_z = \tfrac{1}{4}(1+iz-z^2-iz^3),
    \\
    p_{z^2} = \tfrac{1}{4}(1-z+z^2-z^3), & p_{z^3}=\tfrac{1}{4}(1-iz-z^2+iz^3),
\end{eqnarray}
and the 2-cochain $\omega(g,h)$ whose values are
\begin{equation}
    \omega(z^2,z)=\omega(z^2,z^3)=\omega(z^3,z)=\omega(z^3,z^3)=-1,
\end{equation}
and $\omega(g,h)=+1$ for all other pairs $g,h\in \Z_4$.

The comultiplication stays the same $\Delta^F=\Delta$ since $\C[\Z_4]$ is commutative. The coassociator, on the other hand, becomes
\begin{align}
    \Phi^F_{\C[\Z_4]} &= (1_H\otimes F)(\text{Id}_H\otimes\Delta)(F)(1\otimes 1\otimes 1) (\Delta\otimes\text{Id}_H)(F^{-1})(F^{-1}\otimes1_H),
    \\
    &= 1\otimes 1\otimes 1 -2 (\tfrac{1}{2}(1-z^2))\otimes (\tfrac{1}{2}(1-z^2)) \otimes (\tfrac{1}{2}(1-z^2)).
\end{align}
The linear morphism $\imath: H(2)\to \C^F[\Z_4]$ described by $\imath(1)=1, \imath(g)=z^2$ thus describes the normal inclusion, where in particular $\imath (p_-)= \tfrac{1}{2}(1-z^2)$ and thus
\begin{equation}
    \imath(\Phi_{H(2)}) = \Phi^F_{\C[\Z_4]}.
\end{equation}
Then taking the representation functor $\text{Rep}(-)$ of the exact sequence of quasi-Hopf algebras
\begin{equation}
    H(2) \: \longrightarrow \: \C^F[\Z_4] \: \longrightarrow \: \C[\Z_2]
\end{equation}
gives 
\begin{equation}
    \text{Rep}(\Z_2) \: \cong \: \text{Vec}(\Z_2) \: \longrightarrow \: \text{Rep}(\C^F[\Z_4]) \: \cong \: \text{Vec}(\Z_4) \: \longrightarrow \: \text{Rep}(H(2)) \: \cong \: \text{Vec}(\Z_2,\alpha),
\end{equation}
for $\text{Vec}(\Z_2,\alpha)$ the anomalous $\Z_2$-symmetry.
This describes 
\begin{equation}
    \text{Rep}(\C^F[\Z_4]) \: \cong  \: \text{Vec}(\Z_4)
\end{equation}
as the symmetry category resolving Vec$({\mathbb Z}_2, [\alpha])$.
In addition, we will see in section~\ref{sect:z2:z4:mixedanoom} that the quantum symmetry $\beta$ is encoded in the fact that the functors above are tensor functors, and so also include a set of natural transformations which encode $\beta$.  Altogether, this recovers the picture from section~\ref{sect:ex:anomz2:z4} of an anomalous ${\mathbb Z}_2$ being resolved by a nonanomalous ${\mathbb Z}_4$.

This construction gives us much more than the $\Z_4$ anomaly resolution. For instance, whenever a group $G$ has a normal subgroup $\Z_4$ for which the $\Z_2$ subgroup is central in $G$, then the element (\ref{eq:z4gaugetwist}) is a valid gauge twist for the group algebra $\C[G]$, and thus we have an exact sequence of quasi-Hopf algebras
\begin{equation}\label{eq:resolutionbyz2center}
    H(2)\: \hookrightarrow \: \C^F[Z_4] \: \subset \: \C^F[G] \: \longrightarrow \: \C^F[G]/H(2),
\end{equation}
and thus a non-invertible anomaly resolution
\begin{equation}\label{eq:z2resolutionbyg}
    \text{Rep}(G) \: \longrightarrow \: \text{Vec}(\Z_2,\alpha).
\end{equation}
Some examples of low order are $G=D_4,Q_8$ \cite{Bhardwaj:2024qrf, NS08}.

In the next section, 
we will discuss the mixed anomaly / ``quantum symmetry'' phases in this example.

\subsubsection{Categorical details}

Next, we walk through the details laid out in section~\ref{sect:z2:z4:mixedanoom} for the case of the anomalous ${\mathbb Z}_2$.
Consider a functor $F: {\rm Vec}(\Z_4) \rightarrow {\rm Vec}(\Z_2,\alpha)$, 
where $\Z_4$ has simple objects $g^n$, $n=0,1,2,3$, and $\Z_2$ has simple objects $1$ and $z$.  We will take $\Z_4$ to be non-anomalous, meaning there is a canonical choice of associator with trivial phases.  The $\Z_2$ may have an anomaly, represented by the cocycle $\alpha(z,z,z)=\eta=\pm 1$, and the corresponding nontrivial associator is $a_{z,z,z}=\varphi_z\in{\rm Hom}(z,z)$.  The functor must act on objects as
\begin{equation}
    F(1)=F(g^2)=1,\qquad F(g)=F(g^3)=z,
\end{equation}
and we will take it to act on morphisms simply as $F(\varphi_{g^n})=\varphi_{F(g^n)}=\varphi_{z^n}$.  It remains to determine the strong monoidal structure $J_{g^n,g^m}$, which will be given in terms of a cochain $j$ by
\begin{equation}
    J_{g^n,g^m}=j(g^n,g^m)^{-1}\varphi_{z^{n+m}}.
\end{equation}
The cocycle must satisfy $dj=\pi^\ast\alpha$, or explicitly
(abbreviating $j(g^n,g^m)=j_{n,m}$, with the $n$ and $m$ indices being taken mod 4)
\begin{align}
    \eta^{mnp}=\frac{j_{n,p}j_{m,n+p}}{j_{m,n}j_{m+n,p}},
\end{align}
where we used the fact that $\alpha(z^m,z^n,z^p)=\eta^{mnp}$.  Finding the most general solution to these conditions results in
\begin{equation}
    j_{n,0} \: = \: j_{0,m} \: = \: j_{0,0},
\end{equation}
and
\begin{align}
    j_{3,3}=\ & \eta j_{0,0}j_{1,3}j_{1,2}^{-1},\\
    j_{3,2}=\ & j_{0,0}j_{1,3}j_{1,1}^{-1},\\
    j_{3,1}=\ & \eta j_{1,3},\\
    j_{2,3}=\ & \eta j_{0,0}j_{1,3}j_{1,1}^{-1},\\
    j_{2,2}=\ & j_{1,2}j_{1,3}j_{1,1}^{-1},\\
    j_{2,1}=\ & \eta j_{1,2}.
\end{align}
The phases $j_{0,0}$, $j_{1,1}$, $j_{1,2}$, and $j_{1,3}$ are unfixed by these conditions.  Note that for $n,m>0$, $n\ne m$, we have $j_{n,m}=\eta j_{m,n}$.  This means that the combinations which appear in the $\Z_4$ partition function, namely $j_{n,m}j_{m,n}^{-1}$, are given either by $\eta$ if $n,m>0$ and $n\ne m$, or $1$ otherwise.  That is,
\begin{equation}
    F(Z_{\Z_4})=\frac{1}{4}\sum_{n,m=0}^3j_{n,m}j_{m,n}^{-1}Z_{a^n,a^m}
    \: = \:
    Z_{1,1} + \frac{1+\eta}{2}\lp Z_{1,a}+Z_{a,1}+Z_{a,a}\rp.
\end{equation}
For the case where the $\Z_2$ is non-anomalous, $\eta=1$, this is equal to $2Z_{\Z_2}$, two copies of the $\Z_2$ orbifold, while in the anomalous case, $\eta=-1$, this is $Z_{1,1}=Z$, one copy of the parent theory, in agreement with  \cite{Robbins_2021,Robbins:2021xce,Robbins:2021ibx,Robbins:2022wlr}.

Note that this case can also be formulated in terms of a quantum symmetry phase coming from $H^1(G,H^1(K,U(1)))=H^1(\Z_2,\Z_2)\cong\Z_2$.  In our short exact sequence of groups we can take a section $s(1)=1$, $s(z)=g$.  Then $c(g_1,g_2)=s(g_1)s(g_2)s(g_1g_2)^{-1}$ which defines the extension class obeys $c(1,1)=c(1,z)=c(z,1)=1$, and $c(z,z)=g^2$.  The cohomology group $H^1(G,H^1(K,U(1)))\cong\Z_2$ has one nontrivial element with representative $\beta:G\times K\rightarrow U(1)$ given by $\beta(z^m,g^{2n})=(-1)^{mn}$.  Then $d_2\beta$ is normalized (it equals $1$ if any of its three arguments are $1$) and we have $d_2\beta(z,z,z)=\beta(z,c(z,z))=\beta(z,g^2)=-1$, in other words this $\beta$ maps to the nontrivial anomaly cocycle.  The $j$ that results from this $\beta$ using our formula \eqref{eq:jFromBeta} has nontrivial entries
\begin{equation}
    j_{1,2}=j_{3,2}=j_{1,3}=j_{3,3}=-1,
\end{equation}
corresponding to $j_{0,0}=j_{1,1}=1$, $j_{1,2}=j_{1,3}=\eta=-1$ in the notation above.  This is in agreement with the results of~\cite{Robbins:2021ibx}.

\section{Examples}  \label{sect:exs}

In this section we discuss more general examples of anomaly resolution using noninvertible symmetries.

\subsection{Anomalous $\Z_2$ extended to Rep$(D_4)$}

In this section, we consider again the case of an anomalous ${\mathbb Z}_2$, much as in section~\ref{sect:ex:anomz2:z4:redo}, but instead of extending it to a nonanomalosu ${\mathbb Z}_4$, here we extend to a nonanomalous noninvertible symmetry, following the language of SymTFTs.

To that end, following the general procedure of section~\ref{sect:genl}, we need an example of a non-invertible symmetry which has an igSPT phase.  We saw in section~\ref{sect:ex:anomz2:z4:redo} that the ${\mathbb Z}_4$ resolution could be understood
in terms of an igSPT phase.  Another example with an igSPT phase is Rep$(D_4)$,
see \cite[Table III]{Bhardwaj:2024qrf}.

Here the effectively-acting symmetry is again an anomalous $\Z_2$, which tells us (following
section~\ref{sect:genl}) that one way to resolve an anomalous $\Z_2$ symmetry is via an extension of fusion categories the form
\be
{\rm Vec}(\Z_2\times\Z_2) \: \longrightarrow \: \text{Rep}(D_4) \: \longrightarrow \: 
{\rm Vec}(\Z_2,\alpha).
\ee

Let us first examine how extending a non-anomalous Vec($\Z_2$) to Rep$(D_4)$ works, both from the SymTFT point of view and in concrete partition function computations.  Recall that the partition function for gauging a Rep$(D_4)$ symmetry can be written (given the `usual' choice of gauge for the associator) as
\begin{align}
\label{rd4pf}
Z_{(\gamma_a,\gamma_b,\gamma_c)} \: = \: 
\frac{1}{8}\big[&Z_{1,1}^1 - (Z_{a,b}^c+Z_{a,c}^b+Z_{b,a}^c+Z_{b,c}^a+Z_{c,a}^b+Z_{c,b}^a)
\\\notag
& +(Z_{1,a}^a+Z_{a,1}^a+Z_{a,a}^1)+(Z_{1,b}^b+Z_{b,1}^b+Z_{b,b}^1)
+ (Z_{1,c}^c+Z_{c,1}^c+Z_{c,c}^1)
 \\\notag
& + 2\gamma_a(Z_{1,m}^m+Z_{m,1}^m+Z_{m,m}^1+Z_{a,m}^m+Z_{m,a}^m+Z_{m,m}^a)\\\notag
& + 2\gamma_b(Z_{1,m}^m+Z_{m,1}^m+Z_{m,m}^1+Z_{b,m}^m+Z_{m,b}^m+Z_{m,m}^b)\\\notag
& + 2\gamma_c(Z_{1,m}^m+Z_{m,1}^m+Z_{m,m}^1+Z_{c,m}^m+Z_{m,c}^m+Z_{m,m}^c)\big],
\end{align}
where $(\gamma_a,\gamma_b,\gamma_c)$, which take the values $(-1,1,1)$, $(1,-1,1)$ or $(1,1,-1)$, parameterize the three choices of fiber functor on Rep$(D_4)$, i.e.~the three inequivalent gaugings of the regular representation.

In the SymTFT, for which we will use the notation of \cite[Section II.B]{Bhardwaj:2024qrf} (see also \cite[Appendix A.5]{Iqbal:2023wvm} for additional exposition of the anyons in this model, including their braiding), the initial Rep$(D_4)$-symmetric theory has the algebra $1\oplus e_{RGB}\oplus m_{GB}\oplus m_{RB}\oplus m_{RG}$ on its symmetry boundary -- this trivializes the $D_4$ symmetry on that boundary, leaving a Rep$(D_4)$ which we can take to be given by $\{1,e_{RG},e_R,e_G,m_B\}$.  A theory with an effectively-acting Rep$(D_4)$ symmetry would leave this entire symmetry uncondensed on the physical boundary.  Instead, we would like the $\Z_2\times\Z_2$ subgroup of this Rep$(D_4)$ to be trivially-acting, which we can achieve by taking the algebra 
\be
\label{rd4bphys}
1\oplus e_{RG}\oplus e_R\oplus e_G
\ee
to be the condensible algebra on our physical boundary.

Gauging Rep$(D_4)$ with $(\gamma_a,\gamma_b,\gamma_c)=(-1,1,1)$ corresponds to changing the symmetry boundary condition to $1\oplus e_G\oplus e_R\oplus e_{RG}\oplus 2m_B$; this is the Rep$(D_4)$ gauging `without discrete torsion', i.e.~the one which is dual to gauging the entire $D_4$ symmetry.  The other two gaugings, with $(\gamma_a,\gamma_b,\gamma_c)$ equal to $(1,-1,1)$ and $(1,1,-1)$, correspond to the symmetry boundary conditions $1\oplus e_B\oplus e_G\oplus e_{GB}\oplus 2m_R$ and $1\oplus e_B\oplus e_R\oplus e_{RB}\oplus 2m_G$, respectively.  These are dual to gauging the $\Z_2\times\Z_2$ subgroups of $D_4$ with discrete torsion turned on.  By examining which anyons are condensed on both boundaries we see that, given our choice (\ref{rd4bphys}) for the physical boundary, we expect the $(-1,1,1)$ gauging to produce a theory with four ground states and the other two gaugings to produce theories with two ground states.

We can confirm this by looking at the behavior of the partition function.  Given the trivial action of the $\Z_2\times\Z_2$ subgroup, we expect the partial traces of the Rep$(D_4)$ obifold to map to partial traces of a $\Z_2$ orbifold, as $\Z_2$ is the only effectively-acting symmetry present. Letting $g$ be the generator of that effective $\Z_2$, it must be the case that the elements $\{1,a,b,c\}$ of Rep$(D_4)$ map to the identity in $\Z_2$, with the only non-trivial element coming from $m\rightarrow g+g$.  Letting $i,j,k$ stand for any of $\{a,b,c\}$, the Rep$(D_4)$ traces map to ${\mathbb Z}_2$ orbifold partial traces as follows:
\begin{align}
\label{rd4pt_decomp}
&Z_{1,1}^1\rightarrow Z_{1,1},\hspace{0.5cm}Z_{1,i}^i\rightarrow Z_{1,1},\hspace{0.5cm}Z_{i,1}^i\rightarrow Z_{1,1},\hspace{0.5cm}Z_{i,i}^1\rightarrow Z_{1,1},\hspace{0.5cm}Z_{i,j}^k\rightarrow -Z_{1,1},\\\notag
&Z_{1,m}^m\rightarrow 2Z_{1,g},\hspace{0.5cm}Z_{m,1}^m\rightarrow 2Z_{g,1},\hspace{0.5cm}Z_{m,m}^1\rightarrow 2Z_{g,g},\\\notag
&Z_{a,m}^m\rightarrow -2Z_{1,g},\hspace{0.5cm}Z_{m,a}^m\rightarrow -2Z_{g,1},\hspace{0.5cm}Z_{m,m}^a\rightarrow -2Z_{g,g},\\\notag
&Z_{b,m}^m\rightarrow 2Z_{1,g},\hspace{0.5cm}Z_{m,b}^m\rightarrow 2Z_{g,1},\hspace{0.5cm}Z_{m,m}^b\rightarrow 2Z_{g,g},\\\notag
&Z_{c,m}^m\rightarrow 2Z_{1,g},\hspace{0.5cm}Z_{m,c}^m\rightarrow 2Z_{g,1},\hspace{0.5cm}Z_{m,m}^c\rightarrow 2Z_{g,g}.
\end{align}
The coefficients here can be determined by imposing consistency with the case where the remaining $\Z_2$ acts trivially (hence all of Rep$(D_4)$ acts trivially), and demanding that the result match with \cite[section 5.4.2]{Perez-Lona:2023djo}, \cite{Perez-Lona:2024sds} (after fixing to the appropriate gauge).  Plugging (\ref{rd4pt_decomp}) into (\ref{rd4pf}) produces 
\begin{equation}
    Z_{(\gamma_a,\gamma_b,\gamma_c)} \: = \: 2 Z_{1,1} + \left( \gamma_b + \gamma_c \right)\left( Z_{1,g} + Z_{g,1} + Z_{g,g} \right).
\end{equation}
This result is consisent with the SymTFT calculations: the choice $(\gamma_a,\gamma_b,\gamma_c)=(-1,1,1)$ produces 
\begin{equation}
    Z_{(-1,1,1)} \: = \: 2 \left( Z_{1,1} + Z_{1,g} + Z_{g,1} + Z_{g,g} \right),
\end{equation}
which is the sum of four copies of a $\Z_2$ orbifold partition function.
Similarly, either of $(1,-1,1)$ or $(1,1,-1)$ produces 
\begin{equation}
    Z_{(+1, \mp 1, \pm 1)} \: = \: 2 Z_{1,1},
\end{equation}
the sum of two copies of the parent theory partition function.\footnote{For the latter two cases, the different symmetry boundary algebra will incur additional phases due to braiding, of the type to be discussed below.}\\

Now we turn to the case where the $\Z_2$ symmetry being extended is anomalous.  Here the expectation from the group-like case (particularly from the example of section~\ref{sect:mixedanom:symtft}) is that we should make a different choice of algebra for the symmetry boundary, and in so doing introduce additional phases into the partial trace decompositions.  Indeed, consulting \cite[Table III]{Bhardwaj:2024qrf} suggests that if we want the effective symmetry to be an anomalous $\Z_2$ we should take $1\oplus e_{GB}\oplus e_{RB}\oplus e_{RG}$ as our physical boundary.  Doing so, we find that any of the three gaugings of Rep$(D_4)$ should produce a theory with two ground states.

With such a setup, the TDLs for the trivially-acting $\Z_2\times\Z_2$ on the symmetry boundary are now given by $\{1,e_{GB},e_{RB},e_{RG}\}$ rather than the previous $\{1,e_{RG},e_R,e_G\}$.  This is perhaps clearer when the anyons are labeled in terms of $D_4$ conjugacy classes and irreps of their centralizers.  With the irreps of $D_4$ still denoted $\{1,a,b,c,m\}$, let the order four generator of $D_4$ be $y$ and the order two generator be $x$.  Recall that the conjugacy class $[y^2]$ has centralizer all of $D_4$ and the $[x]$ and $[xy]$ conjugacy classes have $\Z_2\times\Z_2$ centralizers, the $\Z_2\times\Z_2$ irreps for which we label by pluses and minuses.  The anyons we have been working with can then be labeled
\begin{center}
\begin{tabular}{l l l}
$1 \Leftrightarrow (1,1)$,&$e_{RG}\Leftrightarrow (1,a)$,&$e_R\Leftrightarrow(1,b)$\\
$e_G\Leftrightarrow(1,c)$,&$e_{GB}\Leftrightarrow([y^2],b)$,&$e_{RB}\Leftrightarrow([y^2],c),$\\
$m_B\Leftrightarrow(1,m)$,&$m_G\Leftrightarrow([x],+,-)$,&$m_R\Leftrightarrow([xy],+,-)$.
\end{tabular}
\end{center}

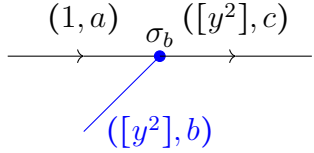
\begin{figure}[H]
	\begin{subfigure}{0.8\textwidth}
		\centering
		\begin{tikzpicture}
        \node at (-1,1.5) {Pure Boundary};
        \draw[thin,->] (0,0) -- (1,0);
        \node at (1,0.5) {$a$};
        \draw[thin] (1,0) -- (2,0);
        \filldraw[black] (2,0) circle (2pt);
        \node at (2,-0.5) {$\sigma_b$};
        \draw[thin,->] (2,0) -- (3,0);
        \node at (3,0.5) {$c$};
        \draw[thin] (3,0) -- (4,0);
        \end{tikzpicture}
        \caption{}
        \label{fig:rd4_diag_a}
    \end{subfigure}
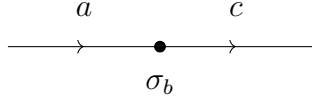
    \begin{subfigure}{0.8\textwidth}
		\centering
		\begin{tikzpicture}
        \node at (0.1,1.5) {Bulk + Boundary, No Anomaly};
        \draw[thin,->] (0,0) -- (1,0);
        \node at (1,0.5) {$(1,a)$};
        \draw[thin] (1,0) -- (2,0);
        \filldraw[blue] (2,0) circle (2pt);
        \node at (2,0.25) {$\sigma_b$};
        \draw[thin,->] (2,0) -- (3,0);
        \node at (3,0.5) {$(1,c)$};
        \draw[thin] (3,0) -- (4,0);
        \draw[blue] (1,-1) -- (2,0);
        \node[blue] at (2,-0.75) {$(1,b)$};
        \end{tikzpicture}
        \caption{}
        \label{fig:rd4_diag_b}
    \end{subfigure}
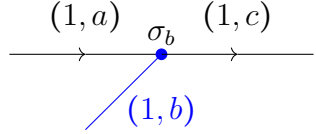
    \begin{subfigure}{0.8\textwidth}
		\centering
		\begin{tikzpicture}
        \node at (0,1.5) {Bulk + Boundary, Anomaly};
        \draw[thin,->] (0,0) -- (1,0);
        \node at (1,0.5) {$(1,a)$};
        \draw[thin] (1,0) -- (2,0);
        \filldraw[blue] (2,0) circle (2pt);
        \node at (2,0.25) {$\sigma_b$};
        \draw[thin,->] (2,0) -- (3,0);
        \node at (3,0.5) {$([y^2],c)$};
        \draw[thin] (3,0) -- (4,0);
        \draw[blue] (1,-1) -- (2,0);
        \node[blue] at (2,-1) {$([y^2],b)$};
        \end{tikzpicture}
        \caption{}
        \label{fig:rd4_diag_c}
    \end{subfigure}
\caption{A diagram of an $a$ line connected to a $c$ line.  There is a topological point operator at their junction, which connects to different bulk anyons depending on whether or not the effective $\Z_2$ symmetry has an anomaly.}
\label{fig:rd4_diag}
\end{figure}

For an example of how this affects the operators appearing in partial traces, see Figure~\ref{fig:rd4_diag}.  Figure~\ref{fig:rd4_diag_a} depicts TDLs which exist in the absolute 2d Rep$(D_4)$-symmetric theory with a trivially-acting $\Z_2\times\Z_2$ subsymmetry, obtained by collapsing the physical and symmetry boundaries onto one another.  The lines labeled by $a$ and $c$ are part of this $\Z_2\times\Z_2$, and its triviality means that they admit a topological junction, the operator for which we label $\sigma_b$.  In the following two figures we imagine the same diagram in which the TDLs are anyons living on the symmetry boundary of a SymTFT.  In Figure~\ref{fig:rd4_diag_b} we assume that the effective $\Z_2$ is non-anomalous, which means that we can take the $\Z_2\times\Z_2$ symmetry to be given by the anyons $\{(1,1),(1,a),(1,b),(1,c)\}$.\footnote{There are multiple choices here for how we could realize the $\Z_2\times\Z_2$ symmetry in terms of anyons, quantified by the nine gSPT phases with reduced topological order $\Z_2$ appearing in \cite[Table III]{Bhardwaj:2024qrf}.  We do not bother examining all of these since the main point is to contrast them with the igSPT phase, which is unique.}  In this case, the operator $\sigma_b$ connects up to a $(1,b)$ line in the bulk.  In Figure~\ref{fig:rd4_diag_c} we instead assume that the effective $\Z_2$ symmetry carries an anomaly, which means that the $\Z_2\times\Z_2$ symmetry must be given by $\{(1,1),(1,a),([y^2],b),([y^2],c)\}$.  Now we see that the anyon which extends into the bulk is $([y^2],b)$.

To see the consequence of this difference on the partial trace decomposition, consider the diagram implied by $Z_{a,m}^m$.  There is an $a$ line which we wish to map to the identity using topological junction operators, akin to the calculations done in Figure~\ref{fig:pt}.  In order to do so, we must pull that topological junction operator past an $m$ line, which has the potential to generate a phase.  We can pinpoint what this phase must be by imagining that the $m$ line in question is raised off of the symmetry boundary, so that instead we are passing the bulk anyon connected to that junction operator through the $m$ line.  For concreteness, assume that when we gauge we take the symmetry boundary to be $1\oplus e_G\oplus e_R\oplus e_{RG}\oplus 2m_B$, which corresponds to $(\gamma_a,\gamma_b,\gamma_c)=(-1,1,1)$.  Then, in decompsing the partial trace $Z_{a,m}^m$ to a multiple of $Z_{1,g}$ we need to pass the $(1,a)=e_{RG}$ anyon through the $(1,m)=m_B$ anyon.  These braid trivially, so we do not pick up any additional phases.  However, in the $Z_{b,m}^m$ diagram we are asked to pass the line $e_{GB}$ through $m_B$, and these in fact have non-trivial braiding, producing a minus sign.  The same goes for all permutations of indices in $Z_{b,m}^m$ and those of $Z_{c,m}^m$ as well (because there we have $e_{RB}$ which braids non-trivially with $m_B$).  In total, the partial trace decomposition we obtain for this case looks like
\begin{align}
\label{rd4pt_decomp_anom1}
&Z_{1,1}^1\rightarrow Z_{1,1},\hspace{0.5cm}Z_{1,i}^i\rightarrow Z_{1,1},\hspace{0.5cm}Z_{i,1}^i\rightarrow Z_{1,1},\hspace{0.5cm}Z_{i,i}^1\rightarrow Z_{1,1},\hspace{0.5cm}Z_{i,j}^k\rightarrow -Z_{1,1},\\\notag
&Z_{1,m}^m\rightarrow 2Z_{1,g},\hspace{0.5cm}Z_{m,1}^m\rightarrow 2Z_{g,1},\hspace{0.5cm}Z_{m,m}^1\rightarrow 2Z_{g,g},\\\notag
&Z_{a,m}^m\rightarrow -2Z_{1,g},\hspace{0.5cm}Z_{m,a}^m\rightarrow -2Z_{g,1},\hspace{0.5cm}Z_{m,m}^a\rightarrow -2Z_{g,g},\\\notag
&Z_{b,m}^m\rightarrow -2Z_{1,g},\hspace{0.5cm}Z_{m,b}^m\rightarrow -2Z_{g,1},\hspace{0.5cm}Z_{m,m}^b\rightarrow -2Z_{g,g},\\\notag
&Z_{c,m}^m\rightarrow -2Z_{1,g},\hspace{0.5cm}Z_{m,c}^m\rightarrow -2Z_{g,1},\hspace{0.5cm}Z_{m,m}^c\rightarrow -2Z_{g,g}.
\end{align}
Plugging these relations into (\ref{rd4pf}) produces $2Z_{1,1}$, i.e. two copies of the parent theory, as we expected to find.

Of course we could have made a different choice for the symmetry boundary, for example $1\oplus e_B\oplus e_G\oplus e_{GB}\oplus 2m_R$, corresponding to $(\gamma_a,\gamma_b,\gamma_c)=(1,-1,1)$.  Now the $m$ line corresponds to the $m_R$ anyon, so we find that the partial traces involving mixes of $a$ with $m$ and $c$ with $m$, which involve crossing $e_{RG}$ and $e_{RB}$ through $m_R$, are the ones that obtain the phases.  Thus the decomposition relations become
\begin{align}
\label{rd4pt_decomp_anom2}
&Z_{1,1}^1\rightarrow Z_{1,1},\hspace{0.5cm}Z_{1,i}^i\rightarrow Z_{1,1},\hspace{0.5cm}Z_{i,1}^i\rightarrow Z_{1,1},\hspace{0.5cm}Z_{i,i}^1\rightarrow Z_{1,1},\hspace{0.5cm}Z_{i,j}^k\rightarrow -Z_{1,1},\\\notag
&Z_{1,m}^m\rightarrow 2Z_{1,g},\hspace{0.5cm}Z_{m,1}^m\rightarrow 2Z_{g,1},\hspace{0.5cm}Z_{m,m}^1\rightarrow 2Z_{g,g},\\\notag
&Z_{a,m}^m\rightarrow 2Z_{1,g},\hspace{0.5cm}Z_{m,a}^m\rightarrow 2Z_{g,1},\hspace{0.5cm}Z_{m,m}^a\rightarrow 2Z_{g,g},\\\notag
&Z_{b,m}^m\rightarrow 2Z_{1,g},\hspace{0.5cm}Z_{m,b}^m\rightarrow 2Z_{g,1},\hspace{0.5cm}Z_{m,m}^b\rightarrow 2Z_{g,g},\\\notag
&Z_{c,m}^m\rightarrow -2Z_{1,g},\hspace{0.5cm}Z_{m,c}^m\rightarrow -2Z_{g,1},\hspace{0.5cm}Z_{m,m}^c\rightarrow -2Z_{g,g}.
\end{align}
Again plugging in recovers two copies of the parent theory.

Finally, for completeness, we could have taken the symmetry algebra $1\oplus e_B\oplus e_R\oplus e_{RB}\oplus 2m_G$.  This gives us $(\gamma_a,\gamma_b,\gamma_c)=(1,1,-1)$ and produces the decomposition relations 
\begin{align}
\label{rd4pt_decomp_anom3}
&Z_{1,1}^1\rightarrow Z_{1,1},\hspace{0.5cm}Z_{1,i}^i\rightarrow Z_{1,1},\hspace{0.5cm}Z_{i,1}^i\rightarrow Z_{1,1},\hspace{0.5cm}Z_{i,i}^1\rightarrow Z_{1,1},\hspace{0.5cm}Z_{i,j}^k\rightarrow -Z_{1,1},\\\notag
&Z_{1,m}^m\rightarrow 2Z_{1,g},\hspace{0.5cm}Z_{m,1}^m\rightarrow 2Z_{g,1},\hspace{0.5cm}Z_{m,m}^1\rightarrow 2Z_{g,g},\\\notag
&Z_{a,m}^m\rightarrow 2Z_{1,g},\hspace{0.5cm}Z_{m,a}^m\rightarrow 2Z_{g,1},\hspace{0.5cm}Z_{m,m}^a\rightarrow 2Z_{g,g},\\\notag
&Z_{b,m}^m\rightarrow -2Z_{1,g},\hspace{0.5cm}Z_{m,b}^m\rightarrow -2Z_{g,1},\hspace{0.5cm}Z_{m,m}^b\rightarrow -2Z_{g,g},\\\notag
&Z_{c,m}^m\rightarrow 2Z_{1,g},\hspace{0.5cm}Z_{m,c}^m\rightarrow 2Z_{g,1},\hspace{0.5cm}Z_{m,m}^c\rightarrow 2Z_{g,g},
\end{align}
again leading to the same result.

In summary, we have seen through both pure SymTFT methods and concrete partition function calculations that when we extend an anomalous $\Z_2$ to a Rep$(D_4)$ symmetry and gauge that symmetry, for any of the three choices of fiber functors we recover two copies of the parent theory and the anomalous $\Z_2$ does not show up in any gaugings, which verifies that the Rep$(D_4)$ is gaugeable/non-anomalous as claimed.  The extra phases which appear in the partition function to make this possible stem from non-trivial anyon braidings in the SymTFT, again matching what we saw in the group-like case.

\subsection{Anomalous $\Z_2$ extended to Rep$(Q_8)$}

We can craft a second example similar to the above one by looking at the representation category of the quaternion group $Q_8$.  We present the eight elements of the group as $\{1,-1,i,-i,j,-j,k,-k\}$ with $\{i,j,k\}$ the order four generators and $\{1,-1\}$ the center.  This group has five irreducible representations, which we will denote $\{1,a,b,c,m\}$.  The irreps $a$, $b$ and $c$ are one-dimensional and have kernels which we will take to be $i$, $j$ and $k$, respectively, while $m$ is the lone two-dimensional irrep.  All together, these five irreps form the same fusion ring as Rep$(D_4)$; it is the associator of Rep$(Q_8)$ which distinguishes it from Rep$(D_4)$ as a fusion category.

The SymTFT for such a symmetry contains operators given by the Drinfeld center $\mcZ(Q_8)$.  As with $D_4$, we can easily write down these anyons which are labeled by a conjugacy class and an irrep of its centralizer.  $Q_8$ has five conjugacy classes:
\be
[1] = \{1\},\hspace{0.3cm}[-1]=\{-1\},\hspace{0.3cm}[i]=\{i,-i\},\hspace{0.3cm}[j]=\{j,-j\},\hspace{0.3cm}[k]=\{k,-k\}.
\ee
The first two of these have the entire group as centralizers, while the latter three have $\Z_4$ centralizers.  This leads to a total of $5+5+4+4+4=22$ anyons, which are summarized in Table~\ref{table:zq8}.  Here we have denoted the irreps of $\Z_4$ as $\{1,i,-1,-i\}$.  We also have indicated the topological spin which describes the lines' self-braiding.  Note that only the 12 bosons, i.e.~anyons with spin 1, can appear in condensable algebras.

\begin{figure}[H]
\begin{center}
\begin{tabular}{|l|l|l|}
\hline
Label & Weight & Spin\\
\hline
$([1],1)$ & 1 & 1\\
$([1],a)$ & 1 & 1\\
$([1],b)$ & 1 & 1\\
$([1],c)$ & 1 & 1\\
$([1],m)$ & 2 & 1\\
$([-1],1)$ & 1 & 1\\
$([-1],a)$ & 1 & 1\\
$([-1],b)$ & 1 & 1\\
$([-1],c)$ & 1 & 1\\
$([-1],m)$ & 2 & -1\\
$([i],1)$ & 2 & 1\\
$([i],i)$ & 2 & i\\
$([i],-1)$ & 2 & -1\\
$([i],-i)$ & 2 & -i\\
$([j],1)$ & 2 & 1\\
$([j],i)$ & 2 & i\\
$([j],-1)$ & 2 & -1\\
$([j],-i)$ & 2 & -i\\
$([k],1)$ & 2 & 1\\
$([k],i)$ & 2 & i\\
$([k],-1)$ & 2 & -1\\
$([k],-i)$ & 2 & -i\\
\hline
\end{tabular}
\end{center}
\caption{The anyons of $\mcZ(Q_8)$.}
\label{table:zq8}
\end{figure}

We can immediately write down the Lagrangian algebra which leads to a Rep$(Q_8)$ symmetry on the symmetry boundary:
\be
\label{rq8_ls}
([1],1)\oplus([-1],1)\oplus([i],1)\oplus([j],1)\oplus([k],1).
\ee
We can see that this has the desired properties, i.e.~it is order 8 and contains five anyons, which matches the number of universes in the Rep$(Q_8)$ SSB phase (which is $Q_8$ gauge theory and therefore contains a universe for each $Q_8$ conjugacy class).  Then, we can identify the algebra which leads to the unique Rep$(Q_8)$ SPT phase as
\be
\label{rq8_spt}
([1],1)\oplus([1],a)\oplus([1],b)\oplus([1],c)\oplus2([1],m).
\ee
This corresponds to the $Q_8$ symmetry on the symmetry boundary.

The remaining Lagrangian algebras, which should fill out the six gaugings of Rep$(Q_8)$, are not hard to suss out.  We should have
\begin{align}
&([1],1)\oplus([1],a)\oplus([-1],1)\oplus([-1],a)\oplus 2([i],1),\\
&([1],1)\oplus([1],b)\oplus([-1],1)\oplus([-1],b)\oplus 2([j],1),\\
&([1],1)\oplus([1],c)\oplus([-1],1)\oplus([-1],c)\oplus 2([k],1)
\end{align}
corresponding to the gauging of the three $\Z_2$ subgroups of Rep$(Q_8)$ and
\be
([1],1)\oplus([1],a)\oplus([1],b)\oplus([1],c)\oplus([-1],1)\oplus([-1],a)\oplus([-1],b)\oplus([-1],c)
\ee
corresponding to gauging the full $\Z_2\times\Z_2$.\\

Now, in order to find a condensable algebra which would lead to an igSPT phase, we require an algebra which has trivial overlap with (\ref{rq8_ls}) and is not a subalgebra of any of the six Lagrangian algebras.  We have three good candidates for such an algebra:
\begin{align}
\label{rq8_anom_a}
&([1],1)\oplus([1],a)\oplus([-1],b)\oplus([-1],c),\\
\label{rq8_anom_b}
&([1],1)\oplus([-1],a)\oplus([1],b)\oplus([-1],c),\\
\label{rq8_anom_c}
&([1],1)\oplus([-1],a)\oplus([-1],b)\oplus([1],c).
\end{align}
(These were discovered by looking for a dimension 4 set of anyons with no mutual braiding; however, not all the axioms have been checked, so we only refer to this as a proposal.)

We can see the effect of putting the three above algebras on the physical boundary at the level of partition functions.  To begin with, we start with the case of a Rep$(Q_8)$-symmetric theory with a trivially-acting $\Z_2\times\Z_2$ subgroup.  This corresponds to putting (\ref{rq8_ls}) on the symmetry boundary and
\be
\label{rq8_z2z2_phys}
([1],1)\oplus([1],a)\oplus([1],b)\oplus([1],c)
\ee
on the physical boundary.  This condenses the $\Z_2\times\Z_2$ subsymmetry of Rep$(Q_8)$, causing it to act trivially in the absolute theory.  The full partition function for gauging Rep$(Q_8)$ is \cite[Section 3.5.3]{Perez-Lona:2023djo}\footnote{Note that the reference contains a sign error in the six member modular orbit.} 
\begin{align}
\label{rq8pf}
\frac{1}{8}\big[&Z_{1,1}^1+(Z_{1,a}^a+Z_{a,1}^a+Z_{a,a}^1)+(Z_{1,b}^b+Z_{b,1}^b+Z_{b,b}^1)+(Z_{1,c}^c+Z_{c,1}^c+Z_{c,c}^1)\\\notag
&-(Z_{a,b}^c+Z_{a,c}^b+Z_{b,a}^c+Z_{b,c}^a+Z_{c,a}^b+Z_{c,b}^a)\\\notag
&+2(Z_{1,m}^m+Z_{m,1}^m+Z_{m,m}^1-Z_{a,m}^m-Z_{m,a}^m-Z_{m,m}^a\\\notag
&-Z_{b,m}^m-Z_{m,b}^m-Z_{m,m}^b-Z_{c,m}^m-Z_{m,c}^m-Z_{m,m}^c)\big].
\end{align}
Identifying the $\Z_2\times\Z_2$ part of this symmetry as trivially-acting gives the following decomposition, where we let $x$, $y$ and $z$ stand in for any of $a$, $b$ and $c$:
\begin{align}
\label{rq8pt_decomp}
&Z_{1,1}^1\rightarrow Z_{1,1},\hspace{0.5cm}Z_{1,x}^x\rightarrow Z_{1,1},\hspace{0.5cm}Z_{x,1}^x\rightarrow Z_{1,1},\hspace{0.5cm}Z_{x,x}^1\rightarrow Z_{1,1},\hspace{0.5cm}Z_{x,y}^z\rightarrow -Z_{1,1},\\\notag
&Z_{1,m}^m\rightarrow 2Z_{1,g},\hspace{0.5cm}Z_{m,1}^m\rightarrow 2Z_{g,1},\hspace{0.5cm}Z_{m,m}^1\rightarrow 2Z_{g,g},\\\notag
&Z_{x,m}^m\rightarrow -2Z_{1,g},\hspace{0.5cm}Z_{m,x}^m\rightarrow -2Z_{g,1},\hspace{0.5cm}Z_{m,m}^x\rightarrow -2Z_{g,g}.
\end{align}
As in the Rep$(D_4)$ case we have fixed the coefficients by demanding consistency with the completely trivially-acting case described in \cite[Section 5.4.3]{Perez-Lona:2023djo}.  Plugging (\ref{rq8pt_decomp}) into (\ref{rq8pf}) produces four copies of a $\Z_2$ orbifold, which matches expectations given the four anyon overlap between (\ref{rq8_spt}) and (\ref{rq8_z2z2_phys}).

Now let us select instead one of the algebras which leads to an igSPT phase; for concreteness, take (\ref{rq8_anom_a}) as the physical boundary algebra.  We should expect to find decomposition relations which mirror (\ref{rq8pt_decomp}) up to modifications from non-trivial anyon braiding.  The trivially-acting $\Z_2\times\Z_2$ symmetry is now given by $\{([1],1),([1],a),([-1],b),([-1],c)\}$, and we see that $([-1],b)$ and $([-1],c)$ have non-trivial braiding with the $m$ line, which is given by $([1],m)$.  In decomposing the partial traces which mix $b$ and $m$ or $c$ and $m$, then, we will acquire additional signs from this braiding, much like the Rep$(D_4)$ case.  The net result is that the decomposition relations (\ref{rq8pt_decomp}) become
\begin{align}
\label{rq8pt_decomp_a}
&Z_{1,1}^1\rightarrow Z_{1,1},\hspace{0.5cm}Z_{1,x}^x\rightarrow Z_{1,1},\hspace{0.5cm}Z_{x,1}^x\rightarrow Z_{1,1},\hspace{0.5cm}Z_{x,x}^1\rightarrow Z_{1,1},\hspace{0.5cm}Z_{x,y}^z\rightarrow -Z_{1,1},\\\notag
&Z_{1,m}^m\rightarrow 2Z_{1,g},\hspace{0.5cm}Z_{m,1}^m\rightarrow 2Z_{g,1},\hspace{0.5cm}Z_{m,m}^1\rightarrow 2Z_{g,g},\\\notag
&Z_{a,m}^m\rightarrow -2Z_{1,g},\hspace{0.5cm}Z_{m,a}^m\rightarrow -2Z_{g,1},\hspace{0.5cm}Z_{m,m}^a\rightarrow -2Z_{g,g},\\\notag
&Z_{b,m}^m\rightarrow 2Z_{1,g},\hspace{0.5cm}Z_{m,b}^m\rightarrow 2Z_{g,1},\hspace{0.5cm}Z_{m,m}^b\rightarrow 2Z_{g,g},\\\notag
&Z_{c,m}^m\rightarrow 2Z_{1,g},\hspace{0.5cm}Z_{m,c}^m\rightarrow 2Z_{g,1},\hspace{0.5cm}Z_{m,m}^c\rightarrow 2Z_{g,g}.
\end{align}
Now, when plugging into (\ref{rq8pf}), the $g$-twisted contributions all cancel and we are left with two copies of the parent theory partition function.  As expected, then, the anomalous $\Z_2$ does not show up in the result, and the full Rep$(Q_8)$ is gaugeable.  If we had instead selected (\ref{rq8_anom_b}) or (\ref{rq8_anom_c}) as our physical boundary condensable algebra, we would have found the same end result, with the signs shuffled to different partial traces.  Note that in each case we expect a decomposition into two universes, given that (\ref{rq8_anom_a}), (\ref{rq8_anom_b}) and (\ref{rq8_anom_c}) have two anyons each overlapping with (\ref{rq8_spt}).  It is also not surprising that we would find three igSPT phases in Rep$(Q_8)$, given that there is an outer automorphism mixing $i$, $j$ and $k$ -- we expect this to translate into an exchange symmetry between $a$, $b$ and $c$.  In $D_4$, by contrast, only $x$ and $xy$ are equivalent by outer automorphism, which leads to asymmetry in Rep$(D_4)$'s own $a$, $b$ and $c$ -- this effect is encoded in the associator.

From the algebraic perspective, following the remark around Equation~(\ref{eq:z2resolutionbyg}) in Section~\ref{sssec:algebraic}, to deduce a $\text{Vec}(\Z_2,\alpha)$ anomaly resolution by $\text{Rep}(Q_8)$ we simply need to find a normal $\Z_4$ subgroup of $Q_8$ whose $\Z_2$ subgroup is central in $Q_8$. In fact, $Q_8$ has \textit{three} different $\Z_4$ normal subgroups satisfying this property. Therefore, one has three different normal inclusions of $H(2)$ into (a gauge twist of) the group algebra $\C Q_8$, so that we have three different anomaly resolutions
\begin{equation}
    \text{Rep}(\Z_2\times \Z_2)\to\text{Rep}(Q_8)\to\text{Vec}(\Z_2,\alpha),
\end{equation}
as suggested by the three candidate condensable algebras proposed above.

\subsection{Further examples: group-theoretic categories}

More examples of anomaly resolutions of invertible symmetries can be obtained via group-theoretical fusion categories. Given a finite group $G$ with a nontrivial 3-cocycle $\omega\in Z^3(G,U(1))$, and an exact factorization $G=HK$ such that $\omega\vert_H\in Z^3(H,U(1))$ is trivial, there is an exact sequence of fusion categories
\begin{equation}
    \text{Rep}(H)\hookrightarrow\mathcal{C}(G,\omega,H,1)\to \text{Vec}(K,\omega\vert_K),
\end{equation}
where $\text{Vec}(K,\omega\vert_K)$ is anomalous due to the nontrivial 3-cocycle $\omega\vert_K\in Z^3(K,U(1))$. This is \cite[Corollary 4.4.i]{Gel17}.  

In the case $\mathcal{C}(G,\omega,H,1)$ is non-anomalous, then this is the anomaly resolution by a $\text{Rep}(H)$ extension (which may involve non-invertible simples).

This recovers well-known results, such as
\begin{equation}
    \text{Rep}(\Z_2\times\Z_2)\hookrightarrow\text{Rep}(D_4)\cong \mathcal{C}(\Z_2^3,\omega,\Z_2^2,1)\to \text{Vec}(\Z_2,\omega\vert_{\Z_2}),
\end{equation}
and
\begin{equation}
    \text{Rep}(\Z_2)\hookrightarrow \text{Rep}(H_8)\cong \mathcal{C}(D_4,\omega,\Z_2,1)\to \text{Vec}(\Z_2\times\Z_2,\omega\vert_{\Z_2\times\Z_2}).
\end{equation}

\section{Conclusions}

In this paper, we have outlined the generalized notion of anomaly resolution in two-dimensional theories from ordinary groups to noninvertible symmetries.  We gave a general discussion in terms of both SymTFT and Hopf algebra constructions, discussed how the ordinary group construction is a special case,
then considered examples in which ordinary groups with anomalies were replaced by noninvertible symmetries.

\section*{Acknowledgements}

We thank Ling Lin, Zhengdi Sun, Yunqin Zheng for valuable discussions. E.S.~and X.Y.~were partially supported by NSF grant
PHY-2310588.

\appendix

\section{Reduced topological order}   \label{app:red}

\subsection{General remarks}
 
 The topological operators appearing the bulk of (d+1)-dimensional SymTFT can capture the generalized charges\cite{Bhardwaj:2023ayw} of a d-dimensional $\mathcal{G}$-Symmetric QFT. Topological operators stretching across both boundaries, after interval compactification produces operators charged under the symmetry. If for some reason such an operator do not `end' on the physical boundary then that particular charged operator will be absent from the absolute theory. So, we can say that the topological operators which do not end along the physical boundary describe the missing charges of the theory\cite{Bhardwaj:2023bbf}. In a situation like this, the symmetry does not act faithfully, only a part of the symmetry acts faithfully, or we can say a part of the symmetry acts trivially.
\begin{figure}[H]
  \centering
    \begin{tikzpicture}[scale=0.75]
    \coordinate (A) at (0,0,0);
    \coordinate (B) at (4,0,0);
    \coordinate (C) at (4,4,0);
    \coordinate (D) at (0,4,0);
    \coordinate (E) at (0,0,4);
    \coordinate (F) at (4,0,4);
    \coordinate (G) at (4,4,4);
    \coordinate (H) at (0,4,4);

    \draw[thick, fill=green!20] (A) -- (B) -- (C) -- (D) -- cycle; 
    \draw[thick, fill=green!20, opacity=0.7] (E) -- (F) -- (G) -- (H) -- cycle; 
    \draw[thick, fill=green!20, opacity=0.7] (A) -- (B) -- (F) -- (E) -- cycle; 
    \draw[thick, fill=blue!20, opacity=0.7] (B) -- (C) -- (G) -- (F) -- cycle; 
    \draw[thick, fill=blue!20, opacity=0.7] (A) -- (D) -- (H) -- (E) -- cycle; 
    \draw[thick, fill=green!20, opacity=0.7] (D) -- (C) -- (G) -- (H) -- cycle; 

    \node at (1.5, 3.5, 1.0) {$\mathfrak{Z}_{d+1}(\mathcal{G})$}; 
    \node at (-0.5, 2.2, 0.5) {$\mathcal{B}^{Sym}_\mathcal{G}$}; 
    \node at (4.5, 3.2, 2.6) {$\mathcal{B}^{Phys}_T$}; 

    \draw[thick, dashed] (-0.5, 1.5, 1.5) -- (3.5, 1.5, 1.5);
    \node at (1.5, 1.8, 1.5) {$\mathcal{Q}_{p+1}$};

    \node at (-0.7, 1.1, 1.5) {$\mathcal{X}_p$};
    \node at (3.7, 1.1, 1.5) {$\mathcal{Y}_p$};

    \fill[black] (-0.5, 1.5, 1.5) circle (2pt);
    \fill[black] (3.5, 1.5, 1.5) circle (2pt);
        \node at (5.5, 1.5, 1.5) {$=$};

        \begin{scope}[shift={(7,0,0)}] 
            \coordinate (P1) at (0,0,0);
            \coordinate (P2) at (0,0,4);
            \coordinate (P3) at (0,4,4);
            \coordinate (P4) at (0,4,0);
            \draw[thick, fill=blue!20, opacity=0.5] (P1) -- (P2) -- (P3) -- (P4) -- cycle; 

            \coordinate (O) at (0, 1.5, 1.5);
            \fill[black] (O) circle (2pt);
            \node at (0.3, 1.8, 1.5) {$\mathcal{O}_p$}; 
        \end{scope}
\end{tikzpicture}
\caption{The $p+1$-dimensional bulk operator ending on the physical boundary along a p-dimensional operator $\mathcal{Y}_p$. The other end of this operator is connected to the symmetry boundary. After interval compactification, we obtain the $\mathcal{G}$-symmetric QFT with an operator $\mathcal{O}_p$, charged under the Symmetry $\mathcal{G}$. If the topological operator $\mathcal{Q}_{p+1}$ does not end along the physical boundary, then that charged operator will be missing from the QFT resulting in a trivially acting symmetry.}
    \label{Missing Charges}
\end{figure}
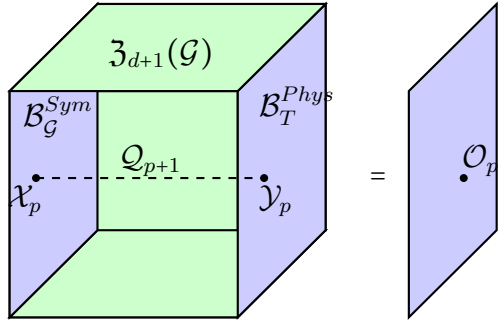
If we have theory with missing charges, we can construct a club sandwich \cite{Bhardwaj:2023bbf,Bhardwaj:2024qrf}, with a topological interface $I_\Phi$ from $\mathfrak{Z}(\mathcal{G}))$ to  $\mathfrak{Z}(\mathcal{G}')$, where, $\mathcal{G}'$ is the effectively acting part of $\mathcal{G}$.
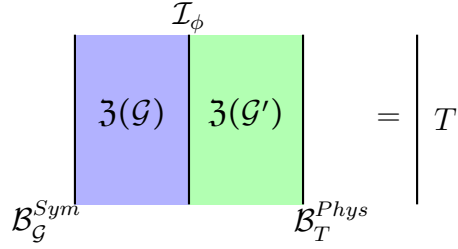
\begin{figure}[H]
    \centering
    \begin{tikzpicture}[scale=0.75]
    
    \fill[blue!30] (0,0) rectangle (2,3); 
    \fill[green!30] (2,0) rectangle (4,3); 
    
    \draw[thick, black] (2,0) -- (2,3);
    
    \draw[thick] (0,0) -- (0,3);
    \draw[thick] (4,0) -- (4,3);
    
    \node at (1.0,1.6) {$\mathfrak{Z}(\mathcal{G})$};
    \node at (3,1.6) {$\mathfrak{Z}(\mathcal{G}')$};

    \node at (-0.5,-0.3) {$\mathcal{B}^{Sym}_\mathcal{G}$}; 
    \node at (4.5,-0.3) {$\mathcal{B}^{Phys}_T$}; 
    \node at (2.0,3.3) {$\mathcal{I}_\phi$}; 

    \node at (5.5,1.5) {=}; 
    \draw[thick, black] (6,0) -- (6,3); 
    \node at (6.5,1.5) {$T$}; 
    \end{tikzpicture}
    \caption{The club sandwich with the respective SymTFT's of $\mathcal{G}$ and $\mathcal{G}'$. Closing the blue side of the club sandwich, we end up with a $\mathcal{G}'$-symmetric boundary condition for $T$. We can write, $\mathcal{B}^{Sym}_{S'} = \mathcal{B}^{Sym}_S \otimes \mathcal{I}_\phi$}
    \label{Club Sandwich}
\end{figure}
\subsection{An explicit example:} 
In this subsection, we would like to identify the `right part' of the club sandwich aka reduced topological order following \cite{Bhardwaj:2023bbf}.
\begin{align*}
    \mathcal{Z}' = \mathcal{Z}/\mathcal{A}
\end{align*}
for the specific case of $\text{Vec}(\Z_4)$. The anyons in the Drinfeld center for $\text{Vec}(\Z_4)$ are given by
\begin{align*}
    \mathcal{Z} = \{e^im^j, \quad i,j= 0,1,2,3\}.
\end{align*}
The Hom space for $\mathcal{Z} =\text{Vec}(\Z_4)$ is 
\begin{align}
    \text{Hom}_\mathcal{Z}(e^im^j,e^km^l) = \begin{dcases}
                                    \C & (i=k,j=l \quad \text{mod} ~ 4), \\
                                    0 & \text{otherwise}.
                                \end{dcases}    
\end{align}
In order to find the consistent fusion algebra $\mathcal{F}$, where the chosen condensable algebra  $\mathcal{A}$ is the vacuum of the condensed theory $\mathcal{F}$. 
\begin{equation}
    \text{Hom}_\mathcal{F}([e^im^j],[e^km^l]) = \text{Hom}_\mathcal{Z}(e^im^j, e^km^l \otimes \mathcal{A}).
\end{equation}
Choosing $\mathcal{A} = \mathcal{A}_e = 1 \oplus e^2$ as our condensable algebra,
\begin{equation}\label{red_top_ord}
\begin{aligned}
    \text{Hom}_\mathcal{F}([e^im^j],[e^km^l]) &= \text{Hom}_\mathcal{Z}(e^im^j, e^km^l \otimes \mathcal{A}_e), \\
    &= \text{Hom}_\mathcal{Z}(e^im^j, e^km^l) \oplus \text{Hom}_\mathcal{Z}(e^im^j,e^{k+2}m^l),
\end{aligned}
\end{equation}
which implies, 
\begin{align}
     \text{Hom}_\mathcal{F}([e^im^j],[e^km^l]) = \begin{dcases}
                                                \C & (i=k \quad \text{mod} ~ 2, j=l \quad \text{mod} ~ 4), \\
                                                 0 & \text{otherwise}.
                                                 \end{dcases}
\end{align}
We can identify $\mathcal{F}$ as, 
\begin{equation}
    \mathcal{F} =\{[e^im^j], \quad i= 0,1 \quad j=0,1,2,3\}.
\end{equation}
In general the simple anyons of $\mathcal{Z}$ will not remain simple in $\mathcal{F}$. We can express a anyon in $\mathcal{Z}$ as superposition of anyons in $\mathcal{F}$. Using the `inverse'/lift map,
\begin{align}
    [e^{i'}m^{j'}] \longrightarrow \bigoplus_{i',j'} n^{i'j'}_{ij} e^im^j
\end{align}
where, 
\begin{align*}
    n^{i'j'}_{ij} = \begin{dcases}
                     1 & \text{if} \quad j=j' \quad \text{mod} ~ 4, i=i' ~ \quad \text{mod}~2, \\
                     0 & \text{otherwise},
                    \end{dcases}                
\end{align*}
replacing $n^{i'j'}_{ij}$, we have the lift
\begin{align}
    [e^{i}m^{j}] \longrightarrow e^im^j + e^{i+2} m^{j}
\end{align}
with the topological spins defined as, 
\begin{align*}
    \theta(e^jm^k) = (i)^{jk}.
\end{align*}
The anyons where every term in the lift has the same spin survive as bulk excitation after condensation, which enforces, 
\begin{align}
    \theta(e^im^j) = \theta(e^{i+2}m^{j}),
\end{align}
so $j$ has to be even, as a result we can identify the reduced topological order to be, 
\begin{align}
    \mathcal{Z}' = \{[e^im^j] \quad i=0,1 \quad j=0,2\}.
\end{align}
Now, moving on to the $\mathcal{A}_{e^2m^2} = 1 \oplus e^2m^2$ condensable algebra, our \eqref{red_top_ord} gets modified, 
\begin{equation}
    \begin{aligned}
    \text{Hom}_\mathcal{F}([e^im^j],[e^km^l]) &= \text{Hom}_\mathcal{Z}(e^im^j, e^km^l \otimes \mathcal{A}_{e^2m^2}), \\
    &= \text{Hom}_\mathcal{Z} (e^im^j, e^km^l) \oplus \text{Hom}_\mathcal{Z}(e^im^j, e^{k+2}m^{l+2} ).
    \end{aligned}
\end{equation}
Once again we identify, 
\begin{align}
     \text{Hom}_\mathcal{F}([e^im^j],[e^km^l]) = \begin{dcases}
                                                \C & (i=k, \quad  j=l \quad \text{mod} ~ 2 \quad \text{and} ~ i+j=k+l \quad \text{mod}~4), \\
                                                 0 & \text{otherwise},
                                                 \end{dcases}
\end{align}
following the same proceedure, 
\begin{align}
    [e^{i}m^{j}] \longrightarrow e^im^j \oplus e^{i+2} m^{j+2}.
\end{align}
The condition from topological spins enforces, $i+j = \text{even}$, that allows us to identify, 
\begin{align}
    \mathcal{Z}' = \{[1],[em],[e^2],[e^3m]\}.
\end{align}
This is the double semion model with $e^2$ being the bosonic $s\bar{s}$.

\section{Condensable algebras}\label{app:condensable}

We review the definition of a condensable algebra in a braided fusion category $\cal D$. See e.g.~\cite{Kong:2013aya} for more.

Condensable algebras are special instances of symmetric special Frobenius algebras. Recall that a symmetric special Frobenius algebra is a tuple $(A,\mu,u,\Delta, u^o)$ where $A\in\text{ob}({\cal D})$ is an object in $\cal D$, and $\mu:A\otimes A\to A$, $u:\mathbbm{1}_{\cal D}\to A$, $\Delta:A\to A\otimes A$, $u^o: A\to \mathbbm{1}_{\cal D}$ are morphisms in $\cal D$ satisfying an array of conditions. We refer the reader to \cite[Appendix A]{Perez-Lona:2023djo} for a complete list of such identities. 

What distinguishes a condensable algebra $(A,\mu,u,\Delta,u^o)$ from a symmetric special Frobenius algebra are the following additional conditions. First, one requires that $A$ is \textit{connected}, meaning that $\text{Hom}(\mathbbm{1}_{\cal D},A)=\C$. Second, one requires it to be \textit{normalized}, so that $u^o\circ u= \text{dim}(A)\,\text{id}_{\mathbbm{1}_{\cal D}}$. Lastly, one requires it to be \textit{(braided-)commutative}. By definition, the braided fusion category $\cal D$ comes equipped with a braiding, a collection of isomorphisms $b_{X,Y}:X\otimes Y\to Y\otimes X$ satisfying some consistency conditions (see e.g. \cite[Chapter 8]{EGNO}). The algebra is called (braided-)commutative if its multiplication morphism $\mu:A\otimes A\to A$ satisfies
\begin{equation}
    \mu = \mu\circ b_{A,A}.
\end{equation}
Note that while the first two conditions can be considered more generally in a fusion category $\cal C$, the third condition only makes sense in a braided fusion category. Thus, a connected normalized commutative symmetric special Frobenius algebra $(A,\mu,u,\Delta,u^o)$ is referred to as a \textit{condensable} algebra.

In the particular case where $\cal D=Z(C)$ is the Drinfeld center of some fusion category $\cal C$, a condensable algebra $A$ satisfying the condition
\begin{equation}
    \vert\text{dim}(A)\vert^2= \text{dim}({\cal Z(C)})
\end{equation}
is known as \textit{Lagrangian} algebra, and specifies a gapped boundary for $\cal Z(C)$.

In physics, a condensable algebra specifies the anyons in the 3d TFT which can end in a two-dimensional boundary.  However, in a two-dimensional theory, to gauge a subsymmetry of a fusion category requires merely a special symmetric Frobenius algebra, not a condensable algebra.

\section{Exact sequence of tensor categories}\label{app:exactseq}

An exact sequence of tensor categories consists of \cite{BN11} a diagram of tensor categories
\begin{equation}\label{eq:appexactsequence}
    {\cal K}\xrightarrow{\imath} {\cal C}\xrightarrow{\pi} {\cal D},
\end{equation}
such that $\pi$ is a \textit{dominant} and \textit{normal} tensor functor, and that the tensor functor $\imath$ is a full embedding whose essential image is ${\cal K}\cong \mathfrak{Ker}_{\pi}\subset {\cal C}$ for $\mathfrak{Ker}_{\pi}$ the kernel of $\pi$, a tensor subcategory of $\cal C$. In particular, this subsumes the notions of exact sequences of groups and Hopf algebras \cite{AD95,Nat20}. Here, by a tensor functor we mean a linear functor equipped with a strong monoidal structure (without braiding) \cite{EGNO}.

We unpack this definition. A \textit{dominant} tensor functor $F:{\cal C}\to {\cal D}$ is a tensor functor such that every object $d\in \text{ob}({\cal D})$ is the subobject of an object $F(c)$ in the image of $F$. That is to say, every object $d$ admits a monomorphism $i:d\to F(c)$ to some $F(c)$.

On the other hand, a tensor functor $F:{\cal C}\to {\cal D}$ is \textit{normal} if for every object $c\in\text{ob}({\cal C})$ there is an object $c_0\in\text{ob}({\cal C})$ such that $F(c_0)$ is the largest trivial subobject of $F(c)$. An object $d$ is called \textit{trivial} if it is isomorphic to a sum of $n$ copies of the monoidal unit $\mathbbm{1}_{\cal D}$ for some $n\in \mathbb{Z}_{\geq 0}$.

Now, given a dominant normal tensor functor $\pi:{\cal C}\to {\cal D}$, one can consider the full subcategory $\mathfrak{Ker}_\pi\subset {\cal C}$ spanned by objects $c\in\text{ob}({\cal C})$ whose image is trivial in $\cal D$. This subcategory is a \textit{tensor} subcategory ${\cal C}$ and is referred to as the \textit{kernel} of $\pi$. In an exact sequence of tensor categories (\ref{eq:appexactsequence}), the image $\imath({\cal K})$ of the embedding category $\imath:{\cal K}\to {\cal C}$ is tensor equivalent $\imath({\cal K})\cong \mathfrak{Ker}_{\pi}$ to the kernel of $\pi$.

\end{document}